\documentclass[11pt,english,twoside]{article}

\usepackage[T1]{fontenc}
\usepackage[latin1]{inputenc}
\usepackage[english]{babel}
\usepackage{lmodern}
\usepackage{a4wide}
\usepackage{amssymb, amsmath, amsthm}
\usepackage{slashed}
\usepackage{float}
\usepackage{graphicx}
\usepackage[dvips]{epsfig}
\usepackage{psfrag}
\usepackage{lscape}
\usepackage[all]{xy}
\usepackage{hyperref}
\usepackage{enumerate}
\usepackage{dsfont}

\usepackage{cite}

\usepackage{mathabx}

\newcommand{\beq}{\begin{equation}}
\newcommand{\eeq}{\end{equation}}
\def\bea#1\eea{\begin{align}#1\end{align}} \newcommand{\nn}{\nonumber}

\newcommand{\id}{\mathds{1}}

\def\del {\partial}
\def\d {{\rm d}}
\def\eee{{\cal E}}
\def\hhh{{\cal H}}

\def\G {\Gamma}
\def\R {\mathcal{R}}
\def\L {\mathcal{L}}
\def\iG {(G^{-1})}
\def\teee{\tilde{{\cal E}}}
\def\tL{\tilde{{\cal L}}}

\def\tg{\tilde{g}}

\def\tR{\widetilde{\mathcal{R}}}
\def\tp {\tilde{\phi}}
\def\b {\beta}
\def\tx{\tilde{x}}
\def\tdel{\tilde{\partial}}
\def\LD {{\cal L}_{{\rm DFT}}}

\newcommand{\be}{\begin{equation}}
\newcommand{\ee}{\end{equation}}

\newcommand{\boxedeq}[1]{
\begin{equation}
\fbox{
\rule[0.7cm]{0pt}{0pt}
$#1$
\rule[-0.45cm]{0pt}{0pt}
}
\end{equation}
}

\makeatletter
\@addtoreset{equation}{section}
\makeatother

\begin{document}

\begin{titlepage}

\rightline{\small LMU-ASC 22/12}
\rightline{\small MPP-2012-67}

\vskip 2.8cm

{\fontsize{18.2}{21}\selectfont
  \flushleft{\noindent\textbf{Non-Geometric Fluxes in Supergravity\\  [0.3cm]
  and Double Field Theory}} }

\vskip 0.2cm
\noindent\rule[1ex]{\textwidth}{1pt}
\vskip 1.3cm

\noindent\textbf{David Andriot$^{a}$, Olaf Hohm$^{a}$, Magdalena Larfors$^{a}$, Dieter L\"ust$^{a,b}$, Peter Patalong$^{a,b}$}

\vskip 0.6cm
\begin{enumerate}[$^a$]
\item \textit{Arnold-Sommerfeld-Center for Theoretical Physics\\Fakult\"at f\"ur Physik, Ludwig-Maximilians-Universit\"at M\"unchen\\Theresienstra\ss e 37, 80333 M\"unchen, Germany}
\vskip 0.2cm
\item \textit{Max-Planck-Institut f\"ur Physik\\F\"ohringer Ring 6, 80805 M\"unchen, Germany}
\end{enumerate}
\noindent {\small{\texttt{david.andriot@physik.uni-muenchen.de, olaf.hohm@physik.uni-muenchen.de,\\magdalena.larfors@physik.uni-muenchen.de, dieter.luest@lmu.de,\\peter.patalong@physik.uni-muenchen.de}}}

\vskip 2.5cm

\begin{center}
{\bf Abstract}
\end{center}

\noindent In this paper we propose ten-dimensional realizations of the non-geometric fluxes $Q$ and $R$. In particular, they appear in the NSNS Lagrangian after performing a field redefinition that takes the form of a T-duality transformation. Double field theory simplifies the 
computation of the field redefinition significantly, and also completes the higher-dimensional picture by providing a geometrical role for the non-geometric fluxes
once the winding derivatives are taken into account. 
The relation to four-dimensional gauged supergravities, together with the global obstructions of non-geometry, are discussed.

\vfill

\end{titlepage}

\tableofcontents

\section{Introduction}\label{sec:intro}

The theory of general relativity provides a
beautiful description of gravity in terms of space-time geometry. According to the
principle of general covariance, the Einstein-Hilbert action is
based on the invariance of the theory under space-time diffeomorphisms, that is, under general coordinate
transformations. 
According to Einstein's field equations, the geometry of space-time is not decoupled from matter, but rather the matter particles back-react when moving in space-time.

In more general terms, the form and even the notion of geometry will  depend on which kind of objects are used to probe space-time. For point particles and their geometrical description one uses differentiable
Riemannian manifolds  that are continuous, and hence the distance between different points on them can be arbitrarily small. 
In string theory there is a lot of convincing evidence that the notion of space-time geometry gets
drastically changed, as compared to that of point particles, when reaching distances that are comparable to the
extension of the string itself. The description of geometry in terms of continuous Riemannian manifolds 
is expected to  break down and to
get replaced
by some `stringy' geometry, which has been thought of in various ways. 
Generically, stringy geometry is characterized by symmetries that have their physical origin in the finite
extent of the string and which suggest an extension of the standard diffeomorphism group of general
relativity. Such a generalization goes beyond standard geometry and is therefore referred to as non-geometry 
\cite{hmw02,dh02,fww04} (see \cite{allp11} for a review on non-geometry).

Mirror symmetry is one well-known example of a stringy symmetry. T-duality is another prime example, which exchanges momentum and winding modes of a closed string on a torus. Since T-duality typically exchanges spaces with large and small radii,
it introduces the notion of a shortest possible distance that can be resolved by a string. More specifically, at large radii the background geometry is probed by the ordinary Kaluza--Klein (KK) momentum modes, and an important part of  their effective, low-energy supergravity action is given by the well-known Neveu--Schwarz (NSNS) Lagrangian 
\beq
{\cal L}= e^{-2\phi} \sqrt{|g|} \left(\R + 4(\partial \phi)^2 - \frac{1}{12} H_{ijk} H^{ijk} \right) . \label{eq:Lns}
\eeq
In the stringy regime, where the radii are of the order of the string length, the mass scale of momentum and winding modes
become comparable, and the effective supergravity description of the momentum modes in general breaks down.

In the simplest case of a constant background metric $g$ and $b$-field, $O(D,D)$ T-duality transformations are just acting as 
automorphisms on the moduli space of string backgrounds.
More generally, one can consider the case of non-constant background fields with non-vanishing NSNS 
$H$ flux. As we will explain in more detail in section \ref{sec:4d},
it has been argued that there exists a chain of T-duality transformations starting with the $H$  flux, leading to four different types of geometrical and non-geometrical fluxes:
\begin{equation}\label{eq:TdualityChain}
H_{abc} \stackrel{T_{a}}{\longrightarrow} f^{a}{}_{bc}
\stackrel{T_{b}}{\longrightarrow} Q_{c}{}^{ab}
\stackrel{T_{c}}{\longrightarrow} R^{abc}\, .
\end{equation}
Here $T_a$ denotes T-dualizing along direction $a$, $H$ is a three-form and $f$ are called geometric fluxes. The latter are given by the first derivatives of the vielbein and are related to the Levi--Civita spin connection and therefore to the curvature of the manifold. On the other hand, the geometric meaning of the $Q$ and $R$ fluxes remains unclear, and will be clarified in this paper.

Originally, the above chain of (non-)geometric fluxes has been discussed in the context of gauged supergravities in lower 
dimensions \cite{stw05,dh05}. These theories, which deform ungauged supergravities by certain mass parameters, can be the result
of flux compactifications of ten-dimensional string theory or supergravity. It turns out, however, that only a subset 
of the consistent gauged supergravities in, say, four dimensions can be obtained through conventional (flux) compactification. 
The four-dimensional scalar potential contains terms corresponding to the geometrical $H$ and $f$ fluxes, 
which have a clear higher-dimensional origin, but also to the $Q$ and $R$ fluxes, which until recently lacked a higher-dimensional 
interpretation. In particular, the gauged supergravities that have a conventional higher-dimensional origin do not appear in 
 a T-duality covariant way, as we will discuss in more detail below.

The main purpose of this paper is to construct a ten-dimensional effective action for the non-geometric $Q$ and $R$ fluxes, which will lead to an understanding of their geometric role. In this we report on results that have recently been announced in  \cite{ahllp12}. At first sight, finding a ten-dimensional realization of $Q$ and $R$ seems to be a difficult task, since they are apparently not part of the NSNS spectrum. In addition, they are thought to correspond to ten-dimensional non-geometric situations. For such configurations, the standard NSNS fields, $g$, $b$ and $\phi$, are not globally well-defined because of the stringy symmetry needed to glue the fields. This prevents a flux compactification to four dimensions.  These two problems were solved in \cite{allp11}, at least for some examples,   using the following field redefinition\footnote{This field redefinition was inspired by studies using generalized geometry, where a relation between $\b$ and non-geometry had been noted \cite{gs06,gs07,gmpw08}. It appeared independently in \cite{hhz10a}.}
\beq
(g_{ij}, b_{ij}, \phi) \, \rightarrow \, (\tg_{ij}, \b^{ij}, \tp) \ ,
\label{eq:fieldredef}
\eeq
 (see \cite{Patalong:2012np,Larfors:2012zz} for summaries of this paper). This field redefinition can be easily characterized in terms of the so-called generalized metric ${\cal H}_{MN}$ for a $D$-dimensional space-time, with $O(D,D)$ indices $M,N=1,\ldots, 2D$, whose inverse can be written as  
\beq
\hhh^{MN}=\begin{pmatrix} g_{ij}-b_{ik} g^{kl}b_{lj} & b_{ik} g^{kj} \\ -g^{ik} b_{kj} & g^{ij}\end{pmatrix} = 
\begin{pmatrix} \tg_{ij} & -\tg_{ik}\b^{kj} \\ \b^{ik} \tg_{kj} & \tg^{ij}-\b^{ik} \tg_{kl} \b^{lj} \end{pmatrix} \ , \label{equality}
\eeq
where $i,j=1,\ldots,D$. We will consider $D=10$ in this paper.

The idea is then that performing the redefinition \eqref{eq:fieldredef} in the NSNS action \eqref{eq:Lns}  makes the non-geometric fluxes appear, in such a way  that the new action with a new metric $\tg_{ij}$, an antisymmetric bi-vector $\b^{ij}$ and a new dilaton $\tp$ is well-defined.
In particular, the $Q$ flux can, at least in some examples, be expressed as the derivative of the bi-vector  $\b^{ij}$ as $Q_k{}^{ij}=\partial_k\b^{ij}$.  In \cite{allp11} a simplifying assumption has been made to the effect that all 
terms involving $\beta^{ij}\partial_{j}$ are ignored, which in turn has the consequence that the $R$ flux disappears. In this paper we relax this assumption and investigate both the ten-dimensional NSNS action and the so-called double field theory (DFT) \cite{hz09a,hz09b,hhz10a,hhz10b} in terms of these new variables.

In ten-dimensional supergravity a fully covariant expression for the $R$ flux can be derived.\footnote{Higher-dimensional expressions for the $Q$ and $R$ fluxes have also been derived in \cite{gmpw08,h09,abmn11,g11,Blumenhagen:2012ma}.} 
We did not find, however, a covariant tensor $Q$ that would reduce to  $Q_k{}^{ij}=\partial_k\b^{ij}$ upon using the assumption, 
and so the role of $Q$ remained somewhat mysterious.  
It turns out that both for the technical problem of writing the action in terms of the new field variables (\ref{eq:fieldredef}) 
and for the geometric interpretation of $Q$ it is of great help to use the formalism of DFT. In particular, the field redefinition (\ref{eq:fieldredef}) takes the form of a T-duality transformation, as we will make more precise below. DFT formally uses  not only the  momentum coordinates $x$, but also the dual winding coordinates $\tilde x$, and hence allows to apply T-duality transformations in absence of isometries. It follows that the complete DFT expression for the $R$ flux also involves derivates with respect to the dual coordinates, as first noted in \cite{abmn11}.
The need for including the dual coordinates in the derivation of the $R$ flux can be understood from the fact that starting from a background with $H$  flux,
the $R$ flux is precisely the complete dual background field, which is seen by the winding modes of the original geometry. The presence of dual derivatives also makes the $R$ flux background not even locally geometric, in accordance with the discussions in \cite{dh05,stw06, hr07}.

Following our recent letter  \cite{ahllp12}, we will provide in this paper a full derivation of an action containing the non-geometric $Q$  and $R$ fluxes. We will show that these fluxes have a particularly nice geometric meaning within DFT, namely as a new connection and a new
covariant tensor. The $Q$ flux arises as a connection rather than a tensor, which allows us to construct a 
derivative for the dual $\tilde{x}$ coordinates that is covariant  \textit{with respect to the $x$ diffeomorphisms}. 
The $R$ flux corresponds to a covariant tensor under $x$ diffeomorphisms, being completely dual to the original $H$  flux. 
Hence the $R$ flux also satisfies a dual Bianchi identity. We are proposing in this way a new geometrical 
calculus that captures aspects of a `stringy' geometry.

The DFT action in terms of the new fields takes the following schematic form:
\bea\label{NewDFTaction}
S_{\rm DFT}(\tilde g,\b,\tilde\phi ) =  \int   dxd\tilde{x}\,\sqrt{|\tg|}\,e^{-2\tilde{\phi}}\Big[&{\cal R} (\tilde g,\partial)
+{\cal R} (\tilde g^{-1},\tilde{\partial})\\ \nonumber
&-\frac{1}{4}Q^2 -\frac{1}{12}R^{ijk}R_{ijk} 
  +4\Big( (\partial\tilde{\phi})^2
  + (\tilde{\partial}\tilde{\phi})^2 \Big) +\dots\Big] 
   \;. 
\eea
There are two Einstein-Hilbert terms: one based on the conventional derivative $\partial_{i}$, and one based
on the winding derivatives $\tilde{\partial}^{i}$, where the inverse metric $\tilde{g}^{ij}$ plays the role of the 
usual metric, and so works consistently with the lower indices of the winding coordinates $\tx_{i}$. 
Even though the first Einstein-Hilbert term is manifestly invariant under $x$ diffeomorphisms $x^{i}\rightarrow x^{i}-\xi^{i}(x)$,
and the second Einstein-Hilbert term is manifestly invariant under $\tilde{x}$ diffeomorphisms $\tilde{x}_{i}\rightarrow \tilde{x}_{i}-\tilde{\xi}_{i}(\tilde{x})$, 
the invariance of the full action as written in (\ref{NewDFTaction}) is not manifest for either of them. 
The reason is that in the full DFT the parameters $\xi^{i}$ and $\tilde{\xi}_{i}$ can a priori depend both on $x$ and $\tilde{x}$. Moreover, 
as mentioned above, $Q$ is not a tensor and therefore the $Q^{2}$ term is not separately diffeomorphism invariant. 
We will show that in our formalism precisely half of the gauge symmetries can be made manifest, here the 
diffeomorphisms parameterized by $\xi^{i}$, by introducing a novel tensor calculus. The $Q$ can then be interpreted as the antisymmetric part of the 
`dual' connection coefficients, so that the $Q^{2}$ term is just part of an extended dual Einstein-Hilbert term. 
In our opinion, this clarifies the geometrical role of the $Q$ flux. 
 
A further aim of this paper  is to relate this action to the known
four-dimensional (gauged) supergravity action after dimensional reduction and thereby to justify our identification 
of $Q$ and $R$ fluxes. 
We will show that the dimensional reduction of the new action (\ref{NewDFTaction}), and in particular its supergravity version, gives precisely rise to 4D potential terms of the expected non-geometric type. 
Our results therefore provide an oxidation of four-dimensional gauged supergravity up to ten dimensions which was previously lacking. This oxidation is not complete, however, as we discuss in detail in section \ref{sec:4d}. The technical reason for this incompleteness is that we only consider supergravity or strongly constrained DFT, whereas some gauged supergravity solutions in dimension $D\ge7$ have been shown to correspond to DFT solutions where the strong constraint is relaxed \cite{Dibitetto:2012rk} (see also \cite{abmn11,g11,gm12}).

We believe, nevertheless, that the new action and field variables considered in this paper provide the first step towards the general case and that 
the novel geometrical structures identified here, or some further generalization, will have to play a role there too. 
Moreover,  it may well be that certain solutions, like de Sitter vacua, that are hidden in the standard 
formulation can be found more easily in the new one, or that global issues that obscure local solutions of the usual action are demystified.     
\medskip 

This paper is organized as follows. In section \ref{sec:DFTrewrite} we provide a short review of DFT and 
write it in terms of the new variables in (\ref{equality}). In section \ref{sec:nongeoflux} we will introduce the
geometrical formalism of the non-geometric fluxes in terms of new connections, covariant derivatives, 
curvature tensors and Bianchi identities with the aim of providing a covariant description of
the theory with respect to standard diffeomorphisms. Section \ref{sec:4d} is devoted to the relation between the 10D and
4D effective actions with non-geometric fluxes, showing in this way that the new formalism provides those missing terms in the 4D action that are required by T-duality covariance.
Computational details are presented in two appendices. In particular, how the field redefinition is performed in DFT is detailed in appendix \ref{ap:calc}, and how it is performed directly in supergravity is shown in appendix \ref{ap:rewrite}.

\newpage
\section{Field redefinition and double field theory}\label{sec:DFTrewrite}

In this section we describe how the action of double field theory (DFT), restricted to the NSNS sector, is rewritten under the field redefinition \eqref{eq:fieldredef} which takes the form of a T-duality. The result is an action that contains an $R$ flux term and several terms that are related to the $Q$ flux. This DFT action can be reduced to a supergravity action and then matches the result obtained after performing the same field redefinition directly in the NSNS action.

\subsection{Generalities of DFT}\label{sec:dftreview}

DFT was introduced in \cite{hz09a,hz09b,hhz10a,hhz10b}, and has been developed in a sequence of papers \cite{Albertsson:2011ux,a11,Copland:2011wx,Copland:2011yh,hk10,hk11,Hohm:2011gs,Hohm:2011dz,Hohm:2011ex,Hohm:2011zr,Hohm:2011dv,Hohm:2011nu,Jeon:2011sq,Jeon:2011vx,Kwak:2010ew,Thompson:2011uw,Schulz:2011ye,Zwiebach:2011rg}. In this theory, T-duality is turned into a manifest symmetry by doubling the coordinates at the level of the effective space-time action for string theory. As already discussed in the introduction, T-duality relates momentum and winding modes of a closed string moving on a torus $T^D$ via the T-duality group $O(D,D)$.\footnote{In general, $D$ refers to the total number of space-time dimensions, that is, here we set
$D=10$. In the context of a KK reduction to $n=D-d$ dimensions on a torus, however, it is more appropriate to double only the $d$ internal coordinates, 
leaving the global $O(d,d)$ T-duality symmetry.}  When the coordinates are doubled, this duality symmetry can be made manifest. Such `doubled' approaches to string theory have previously been studied at the level of the world sheet \cite{d89,Duff:1990hn,t90,t91,h04}, and at the space-time level \cite{t90,t91,h04,s93a,s93b}.

Thus, in DFT every conventional coordinate $x^i$, associated to momentum modes, is complemented by a dual coordinate $\tilde{x}_i$, associated to winding modes. The coordinates combine into a fundamental $O(D,D)$ vector $X^{M}=(\tilde{x}_i,x^{i})$. Although the coordinates are formally doubled we impose  the `strong constraint' 
 \bea
  \eta^{MN}\partial_{M}\partial_{N}  =  0 \;, \qquad 
   \eta^{MN} = \begin{pmatrix} 0 &   1 \\ 1 & 0 \end{pmatrix}\;,
 \eea
where $\eta_{MN}$ denotes the $O(D,D)$ invariant metric and 
$\partial_{M}=(\tilde{\partial}^i,\partial_i)$ denote partial derivatives with respect to the dual coordinates and the standard ones. This constraint is a more stringent version of the level-matching condition in string theory, and is necessary for the consistency of the DFT action, as we will show below. It holds   
on arbitrary fields, parameters and their products, so that 
in particular
 \bea\label{eq:strong}
  \partial_i A\,\tilde{\partial}^i B+  \tilde{\partial}^i  A \, \partial_i B = 0 \;, 
 \eea
for any $A,B$. A consequence of the strong constraint is that the fields locally depend only on half of the coordinates for any DFT solution. 

There are several formulations of DFT. In terms of the generalized metric $\hhh$,  and the dilaton density 
\beq
e^{-2d}=\sqrt{|g|}e^{-2\phi}\ , 
\eeq
where $g={\rm det}g_{ij}$, the DFT action reads \cite{hhz10b}
\bea\label{Hactionx}
\begin{split}
  S_{{\rm DFT}} \ = \ \int dx d\tilde{x}\,e^{-2d}~\Big(~&\frac{1}{8}\,{\cal H}^{MN}\partial_{M}{\cal H}^{KL}
  \,\partial_{N}{\cal H}_{KL}-\frac{1}{2}{\cal H}^{MN}\partial_{N}{\cal H}^{KL}\,\partial_{L}
  {\cal H}_{MK}\\
  &-2\,\partial_{M}d\,\partial_{N}{\cal H}^{MN}+4{\cal H}^{MN}\,\partial_{M}d\,
  \partial_{N}d~\Big)\,.
 \end{split}
 \eea
For our analysis, it is more convenient to express the DFT action in terms of the quantity 
\beq
\eee_{ij}=g_{ij}+b_{ij}\,.
\eeq
Writing out \eqref{Hactionx} in terms of $\eee$, one obtains \cite{hhz10a}
 \bea
\label{eq:sdft}
\begin{split}
  S_{{\rm DFT}} \ = \ \int \,dx d\tilde{x}~
  e^{-2d}\Big[&
  -\frac{1}{4} \,g^{ik}g^{jl}  \,g^{pq} \, \Bigl(
  {\cal D}_{p}{\cal E}_{kl}\,  {\cal D}_{q}{\cal E}_{ij}
  -  {\cal D}_{i}{\cal E}_{lp} {\cal D}_{j}{\cal E}_{kq}
  - \bar{\cal D}_{i}{\cal E}_{pl}\,  \bar{\cal D}_{j}{\cal E}_{qk}
  \Bigr)~
\\ &    + g^{ik}g^{jl}\bigl( {\cal D}_{i} d~
\bar{\cal D}_{j}{\cal E}_{kl}
 +\bar{{\cal D}}_{i} d~{\cal D}_{j}{\cal E}_{lk}\bigr)
 +4g^{ij}{\cal D}_{i}d \,{\cal D}_{j}d ~\Big]\;,\phantom{\biggl(}
 \end{split} 
 \eea
with the calligraphic derivatives  
\begin{equation}\label{eq:calD}
  {\cal D}_{i}   \ = \ \partial_i-{\cal E}_{ik}\tilde{\partial}^k\;, \quad 
    \bar{{\cal D}}_i   \ = \ \partial_i+{\cal E}_{ki}\tilde{\partial}^k\;.
 \end{equation}
Both these DFT actions are background independent.  
 
A crucial property of $S_{{\rm DFT}} $ is that for $\tilde{\partial}^i=0$ it is a rewriting of the standard NSNS action. To be precise, the corresponding DFT Lagrangian $\LD$ reduces to the NSNS Lagrangian up to a total derivative term \cite{hhz10a}
\bea
\label{eq:Ldftns}
\LD\Big|_{\tilde{\partial}=0} \ &= \ e^{-2\phi} \sqrt{|g|} \left(\R + 4(\partial \phi)^2 - \frac{1}{12} H_{ijk}H^{ijk} \right) \\
 &+\ \partial_k \left[ e^{-2\phi} \sqrt{|g|} \left( \partial_m g^{km} + g^{km} g^{ij} \partial_m g_{ij} \right) \right] \ .\nn
\eea
 A second important feature of this DFT action is that it is T-duality invariant under the $O(D,D)$ transformation 
  \begin{equation} \label{fraclin} 
  {\cal E}^{\prime}(X^{\prime}) \ = \ (a{\cal E}(X)+b)(c{\cal E}(X)+d)^{-1}\;,
  \quad
  d^{\prime}(X^{\prime}) \ = \ d(X)\;, \quad X' = h X\,,
 \end{equation}
 where
  \begin{equation}
 h= \begin{pmatrix} a &   b \\ c & d \end{pmatrix} \ \in \ O(D,D)\;.
 \end{equation}
This transformation generalizes the well-known Buscher rules \cite{b87,b88}. More precisely, it has been shown in \cite{hhz10a} that each term of \eqref{eq:sdft} is separately $O(D,D)$ invariant, i.e., 
 \be
  -\frac{1}{4} \,g^{ik}g^{jl}  \,g^{pq} \, 
  {\cal D}_{p}{\cal E}_{kl}\,  {\cal D}_{q}{\cal E}_{ij} \ = \ 
   -\frac{1}{4} \,g^{\prime ik}g^{\prime jl}  \,g^{\prime pq} \, 
  {\cal D}^{\prime}_{p}{\cal E}^{\prime}_{kl}\,  {\cal D}^{\prime}_{q}{\cal E}^{\prime}_{ij}\;, \quad \text{etc.}
 \ee 
These two properties are immensely helpful when performing the field redefinition in DFT, as we will see in the next section.

Finally, DFT also has the essential feature that it is invariant under a generalized diffeomorphism symmetry parametrized by the
$O(D,D)$ vector $\xi^{M}=(\tilde{\xi}_i,\xi^i)$. This symmetry reduces in the supergravity limit $\tilde{\partial}^i=0$ to conventional general coordinate transformations $x^{i}\rightarrow x^{i}-\xi^{i}(x)$ and $b$-field gauge transformations parametrized by $\tilde{\xi}_i$. Conversely, keeping $\tilde{\partial}^i$ non-zero but setting 
$\partial_i=0$, the gauge transformations of DFT reduce in particular to general coordinate transformations in the dual coordinates, $\tilde{x}_i\rightarrow \tilde{x}_i-\tilde{\xi}_i(\tilde{x})$. 
The full gauge symmetry is not manifest in the DFT formulations \eqref{eq:sdft} and \eqref{Hactionx}, but can be verified using the strong constraint \cite{hhz10a}. Employing earlier work by Siegel \cite{s93a,s93b}, more geometrical formulations of DFT have been further developed in \cite{hhz10b,hk10,hz11,jlp10,jlp11}. In these formulations, the DFT action is written in the manifestly gauge invariant form
\beq
S_{\rm DFT} = \int d \tx d x~ e^{-2d}~ {\cal R} ({\hhh}, d) \ ,
\eeq
where ${\cal R} ({\hhh}, d)$ is a generalized curvature scalar associated to $\hhh$ and $d$. For earlier and further interesting work see \cite{Hillmann:2009ci,West:2010ev,Rocen:2010bk,Berman:2010is,Berman:2011pe,Berman:2011kg,Berman:2011cg,Berman:2011jh,Coimbra:2011nw,Coimbra:2011ky}.

\subsection{Field redefinition and T-duality}\label{sec:redef}

In order to identify the ten-dimensional action for $Q$ and $R$ fluxes, we now express the full DFT action \eqref{eq:sdft} in terms of the component fields $\tg, \beta$ and $d$, which are related to $g, b$ and $d$ through the field redefinition \eqref{equality}. It is readily checked, using \eqref{ghat} and \eqref{Bhat} in appendix \ref{ap:rewrite}, that this equality can be rewritten as\footnote{We change conventions with respect to \cite{allp11} so that $\b_{\rm here} = - \b_{\rm there}$.} 
\beq
\label{eq:relation}
(\tg^{-1} + \b)^{-1} \equiv \tilde{\cal E}^{-1} = {\cal E} =  g + b \ ,
\eeq
where we have introduced
\beq
\tilde{\cal E}^{ij} = \tg^{ij} + \b^{ij} \ .
\eeq
We also redefine $\phi$ in order to keep the NSNS measure invariant:
\beq \label{eq:dilintro}
\sqrt{|g|} e^{-2\phi} = e^{-2d} = \sqrt{|\tg|} e^{-2\tp} \ .
\eeq

The redefinition \eqref{eq:relation} has the form of an overall T-duality, which implies that the $O(D,D)$ invariance of DFT renders the associated rewriting particularly simple. Indeed, consider the following $O(D,D)$ transformation, which corresponds to a T-duality in all directions
\be
\label{eq:Tdalldir}
h= \begin{pmatrix} 0 & \id_D \\ \id_D & 0 \end{pmatrix}\quad \Rightarrow \quad x\leftrightarrow \tx \ , \ \del \leftrightarrow \tdel \ , \ \eee(x,\tx) \rightarrow \eee'(x,\tx)=\eee^{-1}(\tx,x) \ , 
\ee
where we have introduced
\be
{\cal E}' =  g' + b' \ .
\ee
Combined with the field redefinition \eqref{eq:relation}, this implies the following set of relations
\be
\label{eq:relprimetilde}
({\cal E}')^{-1}(\tx,x) = {\cal E}(x,\tx) = \tilde{\cal E}^{-1}(x,\tx)  
\ ,
\ee
where the coordinate dependence is written out in order to show that $x\leftrightarrow \tx$ in the T-duality transformation, but not in the field redefinition.\footnote{The position of indices in this relation and \eqref{eq:Tdalldir} can appear confusing. However, 
according to \eqref{fraclin}, we should really write $\eee'(x,\tx)=b\eee^{-1}(\tx,x) c^{-1}$ in \eqref{eq:Tdalldir}, where the matrices $b$ and $c^{-1}$ lower the indices of $\eee^{-1}$ and restore the good index structure. Since for us $b=c=\id_D$, these matrices have been dropped.} Decomposing this equality into its symmetric and antisymmetric parts, we find 
\be
g'(\tx,x) = \tg^{-1} (x,\tx) \ , \ b'(\tx,x)=\b(x,\tx) 
\ .
\ee

Since $\LD$ is $O(D,D)$ invariant, we have that $\LD$ in terms of the T-dual ${\cal E}'$ is the same as the one we started with (in terms of ${\cal E}$). Thus, using this result and \eqref{eq:relprimetilde} we get $\LD$ in terms of the tilded variables without any computation. We simply take $\LD({\cal E}',d)$, and replace 
\beq
\label{eq:caltrans}
\eee_{ij}' \rightarrow \teee^{ij} \ , \quad {\cal D}_{i}'\;\rightarrow \, \tilde{\cal D}^{i} =\tilde{\partial}^{i}-\tilde{\cal E}^{ik}\partial_{k}\;, \quad
  \bar{\cal D}_{i}'\;\rightarrow \, \tilde{\overline{\cal D}}{}^{i} =\tilde{\partial}^{i}+\tilde{\cal E}^{ki}\partial_{k}\; ,
\eeq
which results in the Lagrangian
\bea
\label{eq:Ldfttilde}
\LD(\tg, \b ,d) = e^{-2d}\Big[&
  -\frac{1}{4} \,\tg_{ik}\tg_{jl}\tg_{pq} \, \Bigl(
  \tilde{\cal D}^{p} \teee^{kl}\,  \tilde{\cal D}^{q}\teee^{ij}
  -  \tilde{\cal D}^{i}\teee^{lp} \tilde{\cal D}^{j}\teee^{kq}
  - \tilde{\overline{\cal D}}{}^{i}\teee^{pl}\,  \tilde{\overline{\cal D}}{}^{j}\teee^{qk}
  \Bigr)~ 
\\ &    + \tg_{ik}\tg_{jl}\bigl( \tilde{\cal D}^{i} d~
\tilde{\overline{\cal D}}{}^{j}\teee^{kl}
 +\tilde{\overline{\cal D}}{}^{i} d~\tilde{\cal D}^{j}\teee^{lk}\bigr)
 +4 \tg_{ij}\tilde{\cal D}^{i}d \,\tilde{\cal D}^{j}d ~\Big] \phantom{\biggl(} \ .\nn
 \eea

Before continuing our analysis, let us pause to summarize our method by the diagram
\smallskip
\boxedeq{
 \xymatrix{ \LD({\cal E},d)\ \ar@{=}[r]^{{\rm T-d.}\ {\rm inv.}} & \ \LD({\cal E}',d) \ \ar@{=}[r]^{\eqref{eq:caltrans}} & \ \LD(\tilde{\cal E} ,d) \\
 \L_{{\rm NSNS}}(g,b,d)(x) + \del(\dots)\ \ar@{=}[u]^{\eqref{eq:Ldftns}}_{\tdel=0} \ar@{==}[r] & \ar@{==}[r] &\ \L_{{\rm final}}(\tg,\b,d)(x) + \del(\dots) \ar@{=}[u]_{\tdel=0}  } \label{diagr}
}
\smallskip

\noindent The upper line of this diagram is the rewriting of the DFT Lagrangian just described. The vertical lines correspond to taking the supergravity limit $\tilde{\partial}=0$, and reproduces the NSNS Lagrangian and a Lagrangian written in terms of the new fields, respectively. The dashed equality in the bottom line represents that the field redefinition can be performed directly on the NSNS Lagrangian, as is shown in appendix \ref{ap:rewrite}. Here, we obtain the same equality by going through a chain of DFT Lagrangians, an idea that has been discussed in \cite{allp11}.

When expressing the Lagrangian \eqref{eq:Ldfttilde} in terms of component fields, it is convenient to work with 
 \be
 \label{eq:calD2}
   \tilde{\cal D}^{i} \ = \ -\tilde{g}^{ij}\partial_{j}+\tilde{D}^{i} \;, \qquad 
   \tilde{\overline{\cal D}}{}^i \ = \ \tilde{g}^{ij}\partial_{j}+\tilde{D}^{i} \;, 
  \ee     
where we introduce the derivative operator
\be\label{eq:Dtilde}
  \tilde{D}^{i} \ \equiv \ \tilde{\partial}^i -\beta^{ij}\partial_{j}  \; .
 \ee
A consequence of the strong constraint and the antisymmetry of $\b$ is that, for any fields $A$ and $B$, we have
\beq \label{eq:Dconstraint}
\tilde{D}^i A \partial_i B + \partial_i A \tilde{D}^i B = 0 \ .
\eeq
For now, $\tilde{D}^{i}$ is primarily a convenient book-keeping device, that allows us to separate the conventional Ricci scalar and dilaton term from other terms in the rewritten Lagrangian. However, we will show in section \ref{sec:nongeoflux} that introducing this derivative operator is a first step to a geometric action for the $Q$ and $R$ fluxes.

Re-expressing the DFT Lagrangian in terms of $\tilde{D}^i$ is straightforward, and we refer the reader to appendix \ref{ap:calc} for the explicit calculations. For later convenience we also integrate by parts, thus removing the terms that are linear in dilaton derivatives. After some simplifications the resulting Lagrangian is, up to total derivatives,
 \bea 
e^{2d}  \LD (\tg, \beta, d) &=   {\cal R}(\tg) 
+4(\partial \tp)^2 
+4(\tilde{D} d)^2       
-\frac{1}{4}\tilde{g}_{ik}\tilde{g}_{jl}\tilde{g}_{pq}\big(\tilde{D}^{p}\beta^{kl} \tilde{D}^{q}\beta^{ij}
   -2\tilde{D}^{i}\beta^{lp}\tilde{D}^{j}\beta^{kq}\big) \nn \\
 &-\tilde{D}^i\tilde{D}^j\tilde{g}_{ij}-\frac{1}{4}\tilde{g}_{ik}\tilde{g}_{jl}\tilde{g}_{pq}\big(\tilde{D}^{p}\tilde{g}^{kl} \tilde{D}^{q}\tilde{g}^{ij}
   -2\tilde{D}^{i}\tilde{g}^{lp}\tilde{D}^{j}\tilde{g}^{kq}\big) \label{eq:Ldfttilde1}\\   
   &-2\tilde{g}_{ij}\,\tilde{D}^{i}\partial_{k}{\beta}^{kj}-2\tilde{D}^{i}\tilde{g}_{ij}\, \partial_{k}{\beta}^{kj} 
   -\tilde{g}_{jl}\tilde{g}_{pq}\, \partial_{k}{\beta}^{lp}\tilde{D}^{j}\tilde{g}^{kq} \nn \\
   &-\frac{1}{4}\tilde{g}_{ik}\tilde{g}_{jl}\tilde{g}^{rs}\,\partial_{r}{\beta}^{kl}\,\partial_{s}{\beta}^{ij}-\frac{1}{2}\tilde{g}_{pq}\partial_{k}{\beta}^{lp}\partial_{l}{\beta}^{kq}-\tilde{g}_{ij}\,\partial_{p}{\beta}^{pi}\,\partial_{q}{\beta}^{qj}
 \; . \nn
 \eea

In the above expression we recognize the standard Ricci scalar for the metric $\tg$ in terms of the conventional derivatives $\partial_i$, and the standard kinetic term for the dilaton. The last two terms on the first row in \eqref{eq:Ldfttilde1} are also easily identified; they combine to the square of the $R$ flux
\be\label{eq:rflux}
  R^{ijk}  =  3\tilde{D}^{[i}\beta^{jk]}  \;.
 \ee
This $R$ flux is a tensor, as will be verified in section \ref{sec:nongeoflux}, and represents the covariant field strength of $\beta$. The remaining terms are more difficult to interpret. They contain derivatives of $\beta$ and $\tg$, as we would expect for terms related to the $Q$ flux. However, in contrast to the $R$ flux, it remains obscure how to define a $Q$ flux that reproduces the structures we find. Concretely, it seems difficult to find a tensor whose square gives the relevant terms. We will return to this question, and the geometric interpretation of $Q$, in section \ref{sec:nongeoflux}. Looking ahead to the result we will find, let us relax the covariance condition on $Q$ and define, as in \cite{allp11},
\bea\label{eq:qflux}
  Q_{m}{}^{nk} = \partial_{m}\beta^{nk}\;,
 \eea
so that in the end we have the DFT Lagrangian
 \be\label{eq:Ldfttilde2}
  \begin{split}
e^{2d}  \LD (\tg, \beta, d) &=   {\cal R}(\tg) 
+4(\partial \tp)^2 
+4(\tilde{D} d)^2      
-\frac{1}{12}R^{ijk} R_{ijk}\\
   &-\frac{1}{4}\tilde{g}_{ik}\tilde{g}_{jl}\tilde{g}^{rs}\,Q_{r}{}^{kl}\,Q_{s}{}^{ij}-\frac{1}{2}\tilde{g}_{pq}Q_{k}{}^{lp}Q_{l}{}^{kq}-\tilde{g}_{ij}\,Q_{p}{}^{pi}\,Q_{q}{}^{qj}\\
   &-2\tilde{g}_{ij}\,\tilde{D}^{i}Q_{k}{}^{kj}-2\tilde{D}^{i}\tilde{g}_{ij}\, Q_{k}{}^{kj} 
   -\tilde{g}_{jl}\tilde{g}_{pq}\, Q_{k}{}^{lp}\tilde{D}^{j}\tilde{g}^{kq} \\
 &-\tilde{D}^i\tilde{D}^j\tilde{g}_{ij}-\frac{1}{4}\tilde{g}_{ik}\tilde{g}_{jl}\tilde{g}_{pq}\big(\tilde{D}^{p}\tilde{g}^{kl} \tilde{D}^{q}\tilde{g}^{ij}
   -2\tilde{D}^{i}\tilde{g}^{lp}\tilde{D}^{j}\tilde{g}^{kq}\big)
 \; .
  \end{split}
 \ee

\subsection{Supergravity limit}\label{sec:sugra}

In the supergravity limit the DFT fields are taken to be independent of the dual  coordinates, i.e. one sets $\tilde{\partial}^i = 0$ in the action. This is the final step in the diagram \eqref{diagr}, and allows us to check that the rewriting of the NSNS Lagrangian in appendix \ref{ap:rewrite} agrees with the DFT result. In order to facilitate this check, we use the form \eqref{eq:Ldftcompare} of the DFT Lagrangian, which in the supergravity limit becomes, up to total derivatives,
\bea
e^{2d} \L_{{\rm final}}(\tg,\b,d)(x) &=   {\cal R}(\tg) 
+4(\partial \tp)^2 
  -\frac{1}{12}R^{ijk} R_{ijk} \nn \\
&+4\tilde{g}_{ij}\beta^{ik} \beta^{jl} \partial_k d \, \partial_ld  
-2 \partial_k d \, \partial_l \left(\tg_{ij}\beta^{ik}\beta^{jl} \right)  \nn \\
    &-\frac{1}{4}\tilde{g}_{ik}\tilde{g}_{jl}\tilde{g}^{rs}\,Q_{r}{}^{kl}\,Q_{s}{}^{ij}+\frac{1}{2}\tilde{g}_{pq}Q_{k}{}^{lp}Q_{l}{}^{kq}
    \label{eq:Lfinal} \\  
  &+\tilde{g}_{jl}\tilde{g}_{pq}\beta^{jm}\big(Q_{k}{}^{lp}\partial_m\tilde{g}^{kq}+\partial_{k}\tilde{g}^{lp}\, Q_m{}^{kq}\big) \nn \\
     &-\frac{1}{4}\tilde{g}_{ik}\tilde{g}_{jl}\tilde{g}_{pq}\big(\beta^{pr}\beta^{qs}\partial_r \tilde{g}^{kl} \partial_s\tilde{g}^{ij}
   -2 \beta^{ir}\beta^{js} \partial_r \tilde{g}^{lp}\partial_s\tilde{g}^{kq}\big)\; . \nn
 \eea
Here the $R$ flux term is to be read as the square of the supergravity part of $R$
\be
R|_{\tilde{\partial}=0}^{ijk}  = 3 \beta^{p[i}\partial_{p}\beta^{jk]} \;.
 \ee
 This expression is still covariant. Indeed, using the antisymmetry of the three free indices, and the symmetry of the Christoffel symbols, it can be shown that 
\be\label{covRflux}
R|_{\tilde{\partial}=0}^{ijk}  = 3 \beta^{p[i}\nabla_{p}\beta^{jk]}\;,
 \ee
where $\nabla_{p}$ is the standard covariant derivative, and hence $R|_{\tilde{\partial}=0}$ is a well-defined tensor. Also in the supergravity limit, it is difficult to find a tensor whose square reproduces the last three rows of \eqref{eq:Lfinal}, and so we stick to the definition \eqref{eq:qflux} for the $Q$ flux. In particular, the rather natural guess that $Q$ should be the covariant derivative of $\beta$ does not lead to the observed terms. 

It is readily checked that the Lagrangian \eqref{eq:Lfinal} agrees with the result \eqref{final2} of the rewriting performed in appendix \ref{ap:rewrite}. Furthermore, by comparing \eqref{eq:b5} and \eqref{totder}, and recalling \eqref{eq:Ldftns}, it can be checked that the total derivatives obtained through the two procedures match.

Finally, with $\L_{{\rm final}}$ we have found the generalization of the $Q$ flux Lagrangian that was computed by some of us in \cite{allp11}. Indeed, using a simplifying assumption, \eqref{eq:Lfinal} matches the $Q$ flux Lagrangian found in this paper
\be
e^{2d} \L_{{\rm final}}(\tg,\b,d)(x) \ \
{\buildrel {\beta^{ij} \partial_j = 0 \ , \ \partial_j \beta^{ij}=0} \over \longrightarrow}\ \
  {\cal R}(\tg) 
+4(\partial \tp)^2 
   -\frac{1}{4}\tilde{g}_{ik}\tilde{g}_{jl}\tilde{g}^{rs}\,Q_{r}{}^{kl}\,Q_{s}{}^{ij}\ , \\  
 \ee
and, using the same assumptions, one can also check that the total derivative term matches the one in \cite{allp11}. 

\medskip
In DFT we may equally well solve the strong constraint by setting the conventional derivatives to zero, $\partial_i=0$, 
keeping the winding derivatives $\tilde{\partial}^i$. This corresponds to a T-duality inversion in all directions. The action corresponding to the Lagrangian (\ref{eq:Ldfttilde2}) 
then reduces to 
\be\label{tildeaction}
 S_{\rm DFT} \ = \ \int d\tilde{x}\sqrt{|\det{\tilde{g}^{ij}}|}\,e^{-2\phi^{\prime}}\Big({\cal R}(\tilde{g}^{ij},\tilde{\partial})
 +4\,\tilde{g}_{ij}\,\tilde{\partial}^{i}\phi^{\prime}\,\tilde{\partial}^{j}\phi^{\prime}-\frac{1}{12}R^{ijk}R_{ijk}\Big)  \;,
\ee
where we introduced a new dilaton $\phi^{\prime}$ by 
\be\label{dildef}
   \sqrt{|\det{\tilde{g}^{ij}}|}\,e^{-2\phi^{\prime}} \ = \ e^{-2d}\;, 
\ee
and where the $R$ flux now reads 
\be
 R^{ijk} \ = \ 3\tilde{\partial}^{[i}\beta^{jk]}\;.
\ee
Here, $\tilde{g}^{ij}$ with upper indices plays the role of the metric (rather than the inverse metric) 
on the space with coordinates $\tilde{x}_{i}$. Similarly, this metric appears in the definition (\ref{dildef})
of the new dilaton $\phi^{\prime}$, which guarantees that $\phi^{\prime}$ transforms as a scalar under 
$\tilde{x}$ diffeomorphisms. (In contrast, the dilaton $\tilde{\phi}$ above transforms as a scalar under $x$ diffeomorphisms.) Finally, the field $\beta^{ij}$ transforms as a two-form under  $\tilde{x}$ diffeomorphisms so that 
$R^{ijk}$ plays exactly the same role as the $H$ field strength in the standard NSNS action. 
More generally, as discussed in \cite{hhz10a,Hohm:2011zr}, the whole action (\ref{tildeaction}) is precisely equivalent to 
the NSNS action (\ref{eq:Lns}), just with all upper and lower indices  interchanged and with $(g,b,\phi)$ replaced 
by $(\tilde{g},\beta,\phi^{\prime})$.   Note that the easiest way of obtaining \eqref{tildeaction} is to start from \eqref{conjaction2} and use \eqref{eq:Dphiexpand}.

\section{Geometry of non-geometric fluxes}\label{sec:nongeoflux}
In this section we present a geometrical formalism that allows us to 
write the above DFT action for the new field variables $\tilde{g}_{ij}$ and 
$\beta^{ij}$ in terms of geometrical quantities that make the diffeomorphism 
symmetry in the $x$ coordinates manifest. To this end we introduce novel connections 
that covariantize the winding derivatives $\tilde{\partial}^{i}$ with 
respect to the diffeomorphisms of momentum coordinates, 
and we construct invariant curvatures.

\subsection{Connections for winding derivatives and diffeomorphism invariance}
We begin by recalling the gauge symmetries in DFT spanned by $\xi^{M}=(\tilde{\xi}_i,\xi^{i})$, 
which act on the original field ${\cal E}_{ij}=g_{ij}+b_{ij}$ as  
  \bea\label{gaugeexpand}
  \begin{split}
   \delta {\cal E}_{ij} \ = \ \,&{\cal L}_{\xi}{\cal E}_{ij}
   +\partial_{i}\tilde{\xi}_{j}-\partial_{j}\tilde{\xi}_{i} \\
   &+{\cal L}_{\tilde{\xi}}{\cal E}_{ij}-{\cal E}_{ik}
   \big(\tilde{\partial}^{k}\xi^{l}-\tilde{\partial}^{l}\xi^{k}\big){\cal E}_{lj}\;.
  \end{split}
  \eea
Here, we used the standard Lie derivative with respect to $\xi^{i}$, 
 \be\label{momentumLie}
   {\cal L}_{\xi}{\cal E}_{ij} \ = \    \xi^k\partial_{k}{\cal E}_{ij}
    +\partial_{i}\xi^{k}\hskip1pt{\cal E}_{kj}+\partial_{j}\xi^{k}\,{\cal E}_{ik}\;, 
 \ee
but also a `dual' Lie derivative for winding coordinates with respect to $\tilde{\xi}_i$, 
  \bea\label{dualLie}
  {\cal L}_{\tilde{\xi}}{\cal E}_{ij} \ = 
   \ \tilde{\xi}_{k}\tilde{\partial}^{k}
  {\cal E}_{ij}-\tilde{\partial}^{k}\tilde{\xi}_{i}\,{\cal E}_{kj}
  -\tilde{\partial}^{k}\tilde{\xi}_{j}\,{\cal E}_{ik}\; .
   \eea     
We note that the sign differences between (\ref{momentumLie}) and (\ref{dualLie}) 
reflect the fact that ${\cal E}_{ij}$ is a \textit{covariant} tensor with respect to  
the usual diffeomorphism group but a \textit{contravariant} tensor with respect 
to the dual diffeomorphisms $\tilde{x}_{i}\rightarrow \tilde{x}_{i}-\tilde{\xi}_i$
with lower indices. We infer from (\ref{gaugeexpand}) that the gauge transformation 
parametrized by $\tilde{\xi}_{i}$ has an inhomogeneous term but otherwise acts linearly, 
and that the diffeomorphisms 
parametrized by $\xi^{i}$ act non-linearly. Although not manifest in this form, the 
gauge transformations (\ref{gaugeexpand}) are $O(D,D)$ covariant. 
In particular, the transformation ${\cal E}_{ij}\rightarrow \tilde{\cal E}^{ij}$ 
discussed in sec.~\ref{sec:redef} simply exchanges $\partial_i\rightarrow \tilde{\partial}^i$ and 
$\xi^i\rightarrow \tilde{\xi}_i$,  such that (\ref{gaugeexpand}) becomes 
  \begin{equation}\label{dualvar99}
  \begin{split}
  \delta \tilde{\cal E}^{ij} \ = \ \, & 
   {\cal L}_{\tilde{\xi}}\tilde{\cal E}^{ij} + \tilde{\partial}^{i}\xi^{j}-\tilde{\partial}^{j}\xi^{i}
   \\
  &+ {\cal L}_{\xi}\tilde{\cal E}^{ij}-
  \tilde{\cal E}^{ik}\big(\partial_{k}\tilde{\xi}_{l}-  \partial_{l}\tilde{\xi}_{k}\big)
  \tilde{\cal E}^{lj} \;. 
 \end{split}
 \end{equation}
We observe that in this field basis the $\xi^{i}$ transformations carry an inhomogeneous term but 
are otherwise linear,  
and that the $\tilde{\xi}_i$ transformations are non-linear. In the following we will 
develop a geometrical formalism that renders the  $\xi^{i}$ transformations
manifest, leaving the $\tilde{\xi}_i$ transformations aside for the moment. We will return to  
them in section \ref{invaction}. 
Setting thus $\tilde{\xi}_i=0$ in (\ref{dualvar99}) and decomposing $\tilde{\cal E}^{ij}=\tilde{g}^{ij}+\beta^{ij}$
we obtain 
  \bea\label{transform}
  \delta_{\xi}\tilde{g}_{ij} \ = \ {\cal L}_{\xi}\tilde{g}_{ij}\;, \quad \delta_{\xi}\beta^{ij}
  \ = \  \tilde{\partial}^{i}\xi^{j}-\tilde{\partial}^{j}\xi^{i}+{\cal L}_{\xi}\beta^{ij}\;.
 \eea
We will refer to a transformation under $\xi^{i}$ as covariant if it only involves the Lie derivative
${\cal L}_{\xi}$. In the following we will denote the  
non-covariant part of a variation by $\Delta_{\xi}\equiv \delta_{\xi}-{\cal L}_{\xi}$, 
so that from (\ref{transform})
 \bea\label{Deltagb}
  \Delta_{\xi}\tilde{g}_{ij} \ = \ 0\;, \qquad \Delta_{\xi}\beta^{ij} \ = \  \tilde{\partial}^{i}\xi^{j}-\tilde{\partial}^{j}\xi^{i}\;.
 \eea

Let us now introduce connections for winding derivatives, $\tilde{\partial}^i\rightarrow \tilde{\nabla}^i$, that covariantize 
the `momentum' diffeomorphisms parametrized by $\xi^{i}$. We start by considering a scalar  like 
the dilaton $\tilde{\phi}$, which transforms covariantly, 
 \bea
  \delta_{\xi}\tilde{\phi} \ = \ \xi^{j}\partial_{j}\tilde{\phi}\;.
 \eea 
Therefore, its tilde derivative transforms as 
 \be
 \begin{split}
  \delta_{\xi}(\tilde{\partial}^i\tilde{\phi}) \ = \ \tilde{\partial}^i(\xi^{j}\partial_j\tilde{\phi}) 
  \ = \ \xi^{j}\partial_j(\tilde{\partial}^i\tilde{\phi})+\tilde{\partial}^i \xi^{j}\partial_j\tilde{\phi}\;. 
 \end{split}  
 \ee
Next, we rewrite this transformation in a form that is closer to the Lie derivative 
by adding on the right-hand side 
 \be
  -\partial_j\xi^{i} \tilde{\partial}^j \tilde{\phi}-\tilde{\partial}^j\xi^i \partial_j\tilde{\phi} \ = \ 0\;, 
 \ee
which is zero due to the strong constraint (\ref{eq:strong}), and obtain 
 \be
    \delta_{\xi}(\tilde{\partial}^i\tilde{\phi}) \ = \ {\cal L}_{\xi}(\tilde{\partial}^i\tilde{\phi}) + 
    (\tilde{\partial}^i \xi^{j}-\tilde{\partial}^j\xi^i) \partial_j\tilde{\phi}\;.
 \ee
We infer that the non-covariant term is of the same form as the inhomogeneous 
variation of $\beta$ in (\ref{Deltagb}). Thus, the non-covariant term can be cancelled 
by introducing the derivative operator (\ref{eq:Dtilde}), 
 \be\label{Dtilde}
  \tilde{D}^{i} \ \equiv \  \tilde{\partial}^i -\beta^{ij}\partial_{j}\;,   
 \ee
so that 
 \bea
  \delta_{\xi}(\tilde{D}^i\tilde{\phi}) \ = \ {\cal L}_{\xi}(\tilde{D}^i\tilde{\phi})\;, 
 \eea
and therefore $\Delta_{\xi}(\tilde{D}^i\tilde{\phi}) =0$.     
The derivative (\ref{Dtilde}) will play the role of a partial but 
anholonomic derivative that has a non-trivial commutator, 
    \be\label{commD}
   \big[ \tilde{D}^{i},\tilde{D}^{j}\big] \ = \ -R^{ijk}\partial_{k}-Q_{k}{}^{ij}\tilde{D}^{k}\;, 
  \ee
where as in (\ref{eq:rflux})
 \be\label{Rflux}
  R^{ijk} \ = \ 3\tilde{D}^{[i}\beta^{jk]} \ = \ 3\big(\tilde{\partial}^{[i}\beta^{jk]}+\beta^{p[i}\partial_{p}\beta^{jk]}\big) \;. 
 \ee
The verification of (\ref{commD}) is straightforward but requires the 
strong constraint (\ref{eq:strong}). 

Before we continue with the construction of covariant derivatives we briefly 
discuss that the $R$ flux (\ref{Rflux}) is a covariant tensor under (\ref{transform}). 
In order to see this we first recall that, as noted in (\ref{covRflux}), the second term in $R^{ijk}$ 
can be written in terms of the usual Levi-Civita covariant derivative. This term
is therefore covariant under the  Lie derivative part of the variation (\ref{transform}) of $\beta^{ij}$. 
Since the first term in $R^{ijk}$ takes the form of a curl 
in the winding derivatives it is manifestly invariant under the inhomogeneous 
variation of $\beta^{ij}$ in (\ref{transform}). However, the first term is not covariant 
under the variation by the Lie derivative, and the second term is not covariant 
under the inhomogeneous variation of $\beta^{ij}$, but it turns out that their non-covariant 
variations precisely cancel.  To see this we determine the non-covariant terms in the 
variation of the first term, 
  \be
  \Delta_{\xi}\big(\tilde{\partial}^{[i}\beta^{jk]}\big) \ = \ \tilde{\partial}^{p}\beta^{[ij}\,\partial_{p}\xi^{k]}+
  \partial_p\beta^{[ij}\,\tilde{\partial}^{k]}\xi^{p}+2\tilde{\partial}^{[i}\partial_p\xi^{j}\,\beta^{k]p}\;, 
 \ee 
and the second term, 
 \be
  \Delta_{\xi}\big(\beta^{p[i}\partial_{p}\beta^{jk]}\big) \ = \  
  \partial_{p}\beta^{[ij}\,\tilde{\partial}^{|p|}\xi^{k]}-
  \partial_p\beta^{[ij}\,\tilde{\partial}^{k]}\xi^{p}-2\tilde{\partial}^{[i}\partial_p\xi^{j}\,\beta^{k]p}\;.
 \ee 
The non-covariant variation of $R^{ijk}$ therefore reads
 \be
  \Delta_{\xi}R^{ijk} \ = \  3\big( \tilde{\partial}^{p}\beta^{[ij}\,\partial_{p}\xi^{k]}+\partial_{p}\beta^{[ij}\,\tilde{\partial}^{|p|}\xi^{k]}\big) \ = \ 0\;, 
 \ee
by the strong constraint (\ref{eq:strong}). Thus, $R^{ijk}$ is a covariant tensor 
that can be viewed as the field strength of $\beta^{ij}$.

We now return to the construction of covariant derivatives. For a vector $V^i$ 
and a co-vector $V_{i}$ we set 
 \be\label{covder}
  \tilde{\nabla}^{i}V^{j} \ = \ \tilde{D}^{i}V^{j}-\widecheck{\Gamma}_{k}{}^{ij}V^{k}\,, \quad
  \tilde{\nabla}^{i}V_{j} \ = \ \tilde{D}^{i}V_{j}+\widecheck{\Gamma}_{j}{}^{ik}V_{k}\,, 
 \ee   
which extends in the usual way to tensors with an arbitrary number of upper and lower indices.  
As for the scalar discussed above, $\tilde{D}^{i}V^{j}$ transforms nicely under the transport term, 
but due to the extra term in the Lie derivative of a vector $V^i$ we have 
 \be
  \Delta_{\xi}\big(\tilde{D}^{i}V^{j}\big) \ = \ -\tilde{D}^{i}\partial_k \xi^{j}\,V^{k}\;. 
 \ee
Thus, in order for (\ref{covder}) to transform covariantly, we have to assign the 
following inhomogeneous transformation to the connection components   
 \be\label{simpconn}
  \Delta_{\xi}\widecheck{\Gamma}_{k}{}^{ij} \  = \  -\tilde{D}^{i}\partial_k \xi^{j}\;.
 \ee

Our task is now to determine the connection $\widecheck{\Gamma}_{k}{}^{ij}$ 
in terms of the physical fields. We do so by imposing covariant constraints. 
We first note from (\ref{simpconn}) that the antisymmetric part of $\widecheck{\Gamma}_{k}{}^{ij}$
does not transform as a tensor and therefore cannot be set to zero. 
As the first constraint we demand the usual metricity condition that the metric
is covariantly constant, 
  \be
    \label{metricity}
    \tilde{\nabla}^i\tilde{g}^{jk} \ = \ \tilde{D}^{i}\tilde{g}^{jk}-\widecheck{\Gamma}_{p}{}^{ij}\tilde{g}^{pk}-
    \widecheck{\Gamma}_{p}{}^{ik}\tilde{g}^{jp} \ = \ 0\;.
 \ee  
Since $\widecheck{\Gamma}_{k}{}^{ij}$  has an antisymmetric part, this condition does not 
determine the connection completely, but it allows us to solve for the symmetric part in terms 
of the antisymmetric part and $\tilde{D}^{i}\tilde{g}^{jk}$ in the usual way, 
 \be
  \widecheck{\Gamma}_{k}{}^{(ij)} \ = \ \tilde{\Gamma}_{k}{}^{ij}-\tilde{g}_{kl}\Big(\tilde{g}^{pi}\widecheck{\Gamma}_{p}{}^{[jl]}
  +\tilde{g}^{pj}\widecheck{\Gamma}_{p}{}^{[il]}\Big)\;, 
 \ee
where 
 \be\label{tildeGamma}
  \tilde{\Gamma}_{k}{}^{ij} \ = \  \frac{1}{2}\tilde{g}_{kl}\left(\tilde{D}^i\tilde{g}^{jl}+  \tilde{D}^j\tilde{g}^{il}-\tilde{D}^l\tilde{g}^{ij}\right)\;
 \ee    
are the conventional Christoffel symbols in the winding coordinates, 
with $\tilde{\partial}^i$ replaced by $\tilde{D}^i$.  
In order to determine the antisymmetric part we consider the 
commutator of covariant derivatives on a scalar $\tilde{\phi}$, 
  \bea\label{tildeDcomm}
  \begin{split}
   \big[ \tilde{\nabla}^i,\tilde{\nabla}^j\big]\tilde{\phi} \ &= \ 
   \big[ \tilde{D}^{i},\tilde{D}^{j}\big]\tilde{\phi}-\widecheck{\Gamma}_{k}{}^{ij}\tilde{D}^{k}\tilde{\phi}+\widecheck{\Gamma}_{k}{}^{ji}\tilde{D}^{k}\tilde{\phi} \\
   \ &= \ 
   -R^{ijk}\partial_k\tilde{\phi} -\big(Q_{k}{}^{ij}+2\widecheck{\Gamma}_{k}{}^{[ij]}\big)\tilde{D}^{k}\tilde{\phi}\;, 
  \end{split} 
 \eea
where we used (\ref{commD}). As $R^{ijk}$ transforms as a tensor it is a covariant 
condition to demand that the commutator (\ref{tildeDcomm}) is given by the $R$ flux only, 
 \bea\label{commconstr}
  \big[ \tilde{\nabla}^i,\tilde{\nabla}^j\big]\tilde{\phi} \ =  \ -R^{ijk}\partial_k\tilde{\phi} \;.
 \eea
This constraint is then solved by    
 \bea\label{GammaQ}
  \widecheck{\Gamma}_{k}{}^{[ij]} \ = \ -\frac{1}{2}Q_{k}{}^{ij}\;.
 \eea
In total, the covariant constraints (\ref{metricity}) and (\ref{commconstr}) determine 
the connection completely, which is given by 
 \bea\label{fullconn}
  \widecheck{\Gamma}_{k}{}^{ij} \ = \ \tilde{\Gamma}_{k}{}^{ij}+\tilde{g}_{kl}\tilde{g}^{p(i}Q_{p}{}^{j)l}-\frac{1}{2}Q_{k}{}^{ij}\;. 
 \eea
By construction, this transforms as required by (\ref{simpconn}), 
which one may also verify explicitly.

\subsection{Bianchi identities and invariant  curvatures}
After having defined connections and covariant derivatives we apply 
them now by discussing Bianchi identities and invariant curvatures. 
We start by noting that the $R$ flux satisfies the Bianchi identity 
  \be\label{RfluxBianchi}
   \tilde{\nabla}^{[i}R^{jkl]} \ = \ 0\;, 
  \ee
which reads explicitly 
  \be
  4\,\tilde{\partial}^{[i}R^{jkl]}+4\,\beta^{p[i}\partial_{p}R^{jkl]}+6\,Q_{p}{}^{[ij} R^{kl]p} \ = \ 0\;. 
 \ee
In this form the Bianchi identity can be verified by a straightforward 
computation.

We now construct invariant curvatures. In addition to a Riemann curvature tensor
that appears through the commutator of covariant derivatives, using the familiar 
$[\partial_i,\partial_j]=0$, there is a new torsion due to the new constraint $\tilde{\partial}^i\partial_i=0$.
We find 
 \bea
  \tilde{\nabla}^{i}\nabla_{i}\tilde{\phi} \ = \ \nabla_{i}\tilde{\nabla}^{i}\tilde{\phi} \ = \ {\cal T}^{i}\nabla_{i}\tilde{\phi}\;,
 \eea
 where   
 \be\label{Gammatrace}
  {\cal T}^{i} \ = \  \widecheck{\Gamma}_{k}{}^{ki} \ = \ \frac{1}{2}\tilde{g}_{pq}\tilde{D}^i \tilde{g}^{pq}-Q_{k}{}^{ki}\;.
 \ee
Thus, curiously, the trace of the connection transforms as a tensor, which can 
also be verified directly with the transformation rule (\ref{simpconn}), 
 \be
  \Delta_{\xi}\widecheck{\Gamma}_{k}{}^{kj} \ = \ -\partial_{k}\tilde{\partial}^{k}\xi^{j}+\beta^{kp}\partial_p \partial_k\xi^{j} \ = \ 0\;, 
 \ee
using the strong constraint and the antisymmetry of $\beta$. Thus, while the usual torsion tensor given 
by the antisymmetric part of the connection coefficients does not have tensor character in this formalism, 
as noted above, the trace of the connection is a tensor and therefore naturally viewed as a new torsion. 

We note that the rules for partial integration involve 
the new torsion ${\cal T}^{i}$:
  \be\label{piconstr}
  \int dxd\tilde{x} \sqrt{|\tg|}\,V_i\tilde{\nabla}^i W \ = \  
    -\int dxd\tilde{x} \sqrt{|\tg|}\,W\big(\tilde{\nabla}^i V_i-2{\cal T}^{i}V_{i}\big)\;, 
 \ee
as may be verified with (\ref{Gammatrace}).    

Next, we define a covariant curvature or Riemann tensor and 
prove algebraic and differential Bianchi identities. 
A Riemann tensor can be defined through the commutator of 
covariant derivatives on a co-vector, 
  \be\label{fullcomm}
   \big[ \tilde{\nabla}^{i},\tilde{\nabla}^{j}\big]V_{k} \ = \ -R^{ijp}\nabla_{p}V_{k}+ \widecheck{\cal R}^{ij}{}_{k}{}^{l}V_{l}\;, 
 \ee 
where 
 \be\label{RIEMANN}
   \widecheck{\cal R}^{ij}{}_{k}{}^{l} \ =  \ \tilde{D}^{i}\widecheck{\Gamma}_k{}^{jl}-\tilde{D}^{j}\widecheck{\Gamma}_{k}{}^{il}
  +\widecheck{\Gamma}_{k}{}^{iq}\,\widecheck{\Gamma}_{q}{}^{jl}- \widecheck{\Gamma}_{k}{}^{jq}\,\widecheck{\Gamma}_{q}{}^{il} 
  +Q_{q}{}^{ij}\,\widecheck{\Gamma}_k{}^{ql}-R^{ijq}\,\Gamma^{l}{}_{qk}\;.
 \ee   
The verification of (\ref{fullcomm}) requires (\ref{commD}). Note the appearance 
of the conventional Christoffel symbols $\Gamma^{k}{}_{ij}$ (with respect to $\tilde{g}_{ij}$) in the definition of the Riemann tensor. 
In (\ref{fullcomm}) this term cancels against the connection inside the covariant derivative in the 
first term. We have written this first term with a covariant derivative 
such that it is separately diffeomorphism invariant. Therefore, with the left hand side of (\ref{fullcomm})
being manifestly covariant, the Riemann tensor (\ref{RIEMANN}) must be a covariant tensor 
as well, as one may also verify explicitly.    
The commutator of covariant derivatives on a vector reads similarly 
  \be\label{fullcomm2}
   \big[ \tilde{\nabla}^{i},\tilde{\nabla}^{j}\big]V^{k} \ = \ -R^{ijp}\nabla_{p}V^{k}- \widecheck{\cal R}^{ij}{}_{l}{}^{k}V^{l}\;. 
 \ee

Let us now discuss the symmetry properties of the Riemann tensor. By construction 
it is manifestly antisymmetric in its first two indices. As a consequence of the 
metricity condition it is also antisymmetric in its last two indices, which can be 
proved as follows. Raising the index $k$ on both sides of (\ref{fullcomm}) 
we obtain 
 \be
  \big[ \tilde{\nabla}^{i},\tilde{\nabla}^{j}\big]V^{k} \ = \ -R^{ijp}\nabla_{p}V^{k}+ \widecheck{\cal R}^{ijkl}V_{l}\;, 
 \ee 
which by comparison with (\ref{fullcomm2}) implies 
 \be
  \widecheck{\cal R}^{ijkl} \ = \ -\widecheck{\cal R}^{ijlk}\;.
 \ee
There is no exchange symmetry between the two index pairs, because 
this Riemann tensor satisfies a modified Bianchi identity. 
We derive this Bianchi identity from the Jacobi identity 
 \be\label{Jacobi}
  \Big(\big[\tilde{\nabla}^{i},\big[\tilde{\nabla}^{j},\tilde{\nabla}^{k}\big]\big]
  +\big[\tilde{\nabla}^{j},\big[\tilde{\nabla}^{k},\tilde{\nabla}^{i}\big]\big]
  +\big[\tilde{\nabla}^{k},\big[\tilde{\nabla}^{i},\tilde{\nabla}^{j}\big]\big]\Big)\tilde{\phi} \ = \ 0\;.
\ee
Using (\ref{commconstr}) and (\ref{fullcomm2}) this implies 
 \be
  0 \ = \ -\tilde{\nabla}^{[i}R^{jk]l}\,\partial_{l}\tilde{\phi}-R^{l[ij}\,\big[\tilde{\nabla}^{k]},\nabla_{l}\big]\tilde{\phi}
  +\widecheck{\cal R}^{[ij}{}_{l}{}^{k]}\,\tilde{\nabla}^{l}\tilde{\phi}\;.
 \ee
This can be simplified using 
 \be
  \big[\tilde{\nabla}^{i},\nabla_l\big]\tilde{\phi} \ = \ \widecheck{\Gamma}_{l}{}^{pi}\partial_{p}\tilde{\phi}-\Gamma^{i}{}_{lp}\tilde{D}^{p}\tilde{\phi}\;.
 \ee
We finally get 
 \be
  0 \ = \ -4\tilde{\nabla}^{[i}R^{jkl]}\,\nabla_{l}\tilde{\phi}+3\widecheck{\cal R}^{[ij}{}_{l}{}^{k]}\,\tilde{\nabla}^{l}\tilde{\phi}+
  \nabla_{p}R^{ijk}\,\tilde{D}^{p}\tilde{\phi}\;, 
 \ee
where we used that the strong constraint  implies 
 \be
  \tilde{D}^{l}R^{ijk}\,\partial_{l}\tilde{\phi}+\partial_{l}R^{ijk}\,\tilde{D}^{l}\tilde{\phi} \ = \ 0 \;.
 \ee
Therefore, using the Bianchi identity (\ref{RfluxBianchi}) for the $R$ flux, 
we obtain the algebraic Bianchi identity  
 \be\label{algBianchi}
   3\widecheck{\cal R}^{[ij}{}_{l}{}^{k]} +
  \nabla_{l}R^{ijk}  \ = \ 0\;.
 \ee      
After raising the index $l$ this identity reads explicitly
 \bea
  \widecheck{\cal R}^{ijkl}+\widecheck{\cal R}^{jkil}+\widecheck{\cal R}^{kijl} \ = \ \nabla^{l}R^{ijk} \;.
 \eea 
 Writing this equation with four different index permutations and taking an appropriate 
 linear combination it is straightforward to derive  
 \be\label{newexchange}       
  \widecheck{\cal R}^{ijkl}-\widecheck{\cal R}^{klij} \ = \ \nabla^{[i}R^{j]kl}-\nabla^{[k}R^{l]ij}\;.
 \ee
Thus, under exchange of the index pairs the Riemann tensor goes into itself only 
up to corrections involving the covariant derivative of the $R$ flux. We finally 
give an alternative, more explicit form of (\ref{algBianchi}) by writing out the 
connections, 
 \be
  \partial_l R^{ijk} \ = \ 3\big(\tilde{D}^{[i}Q_{l}{}^{jk]}-Q_{q}{}^{[ij}Q_{l}{}^{k]q}\big)\;.
 \ee
Upon setting $\tilde{\partial}^i=0$ this reduces to eq.~(75) in \cite{Blumenhagen:2012ma}.

\medskip

Let us now introduce a Ricci tensor and Ricci scalar. Due to the antisymmetry in 
each index pair of the Riemann tensor there is one independent non-vanishing contraction,  
 \be
  \widecheck{\cal R}^{ij} \ \equiv \ \widecheck{\cal R}^{ki}{}_{k}{}^{j} \;.
 \ee   
Explicitly this reads  
 \be\label{finalRicci}
    \widecheck{\cal R}^{ij}  \ = \ \tilde{D}^{k}\widecheck{\Gamma}_{k}{}^{ij}-\tilde{D}^{i}\widecheck{\Gamma}_{k}{}^{kj}+
    \widecheck{\Gamma}_{k}{}^{ij}\,\widecheck{\Gamma}_{q}{}^{qk}-\widecheck{\Gamma}_{p}{}^{ki}\,\widecheck{\Gamma}_{k}{}^{pj}  \;, 
 \ee    
where we used (\ref{GammaQ}) to simplify and combine terms.  
Recalling that the trace of the connection equals the tensor (\ref{Gammatrace}) this can also be written 
as 
 \be
    \widecheck{\cal R}^{ij}  \ = \ \tilde{D}^{k}\widecheck{\Gamma}_{k}{}^{ij}
    -\widecheck{\Gamma}_{q}{}^{ki}\,\widecheck{\Gamma}_{k}{}^{qj}-\tilde{\nabla}^{i}{\cal T}^{j}\;.
 \ee
Remarkably, the Ricci tensor thus decomposes into two structures that have separately tensor 
character.  The Ricci tensor is not symmetric, rather by taking the trace of (\ref{newexchange}) we 
infer 
 \bea
  \widecheck{\cal R}^{[ij]} \ = \ -\frac{1}{2}\nabla_{k}R^{kij}\;, 
 \eea
i.e., the antisymmetric part of the Ricci tensor is determined by the $R$ flux.   
Finally we can define a scalar curvature in the usual way, 
  \bea\label{newscalar}
   \widecheck{\cal R} \ = \ \tilde{g}_{ij}\widecheck{\cal R}^{ij}\;, 
  \eea
which by construction is a scalar under diffeomorphisms and can thus 
be used to define an invariant action.

\subsection{Invariant action}\label{invaction}
We have now all ingredients at hand in order to write the full DFT action 
in terms of the geometrical objects introduced above, 
  \be\label{conjaction}
 \begin{split}
  S_{\rm DFT} \ = \ \int   dxd\tilde{x}\,\sqrt{|\tg|}\,e^{-2\tilde{\phi}}\Big[&{\cal R} + \widecheck{\cal R}-\frac{1}{12}R^{ijk} R_{ijk} \\
  &
  +4\Big( (\partial\tilde{\phi})^2
  + (\tilde{D}\tilde{\phi})^2 
  +\tilde{\nabla}^{i}{\cal T}_{i}-{\cal T}^{i}{\cal T}_{i}\Big) \Big] 
   \;. 
 \end{split}
 \ee     
In here, every term is manifestly diffeomorphism invariant. This action contains two 
Einstein-Hilbert terms. The first is the conventional one based on derivatives $\partial_i$
and the metric $\tilde{g}_{ij}$. The second one is based on the novel Ricci scalar  (\ref{newscalar})
that involves winding derivatives and generalized connections that contain the $Q$ flux 
as the antisymmetric part.  Moreover, the new torsion ${\cal T}^{i}$ is required. 
It is shown in appendix \ref{ap:calc} that this action indeed equals the DFT action in terms 
of $\tilde{g}_{ij}$, $\beta^{ij}$ and $\tilde{\phi}$ determined in \eqref{eq:Ldfttilde2}.  

Up to total derivatives the terms involving the new torsion ${\cal T}^i$ can be rewritten 
as a square with the $\tilde{D}^i\tilde{\phi}$ terms as follows. We have by (\ref{piconstr}) 
 \be
  4\int dxd\tilde{x}\sqrt{|\tg|}e^{-2\tilde{\phi}}\tilde{\nabla}^{i}{\cal T}_{i}
  \ = \ 4\int dxd\tilde{x}\sqrt{|\tg|}e^{-2\tilde{\phi}}\big(2\tilde{\nabla}^{i}\tilde{\phi}\,{\cal T}_{i}
  +2{\cal T}^{i}{\cal T}_{i}\big)\;, 
 \ee
and therefore 
 \be
   4\int dxd\tilde{x}\sqrt{|\tg|}e^{-2\tilde{\phi}}\big( \tilde{\nabla}^{i}{\cal T}_{i}-{\cal T}^{i}{\cal T}_{i}\big)
   \ = \   4\int dxd\tilde{x}\sqrt{|\tg|}e^{-2\tilde{\phi}}\big(2\tilde{\nabla}^{i}\tilde{\phi}\,{\cal T}_{i}
  +{\cal T}^{i}{\cal T}_{i}\big)\;.
 \ee
The full action (\ref{conjaction}) can then be written as 
  \be\label{conjaction2}
  S_{\rm DFT} \ = \ \int   dxd\tilde{x}\,\sqrt{|\tg|}\,e^{-2\tilde{\phi}}\Big[{\cal R} + \widecheck{\cal R}-\frac{1}{12}R^{ijk} R_{ijk} 
  +4 (\partial \tilde{\phi})^2
  + 4\big(\tilde{D}^i\tilde{\phi}+{\cal T}^i\big)^2  \Big]
   \;. 
 \ee     

Let us note that this action is particularly convenient to derive \eqref{tildeaction}, where $\widecheck{\cal R}$ reduces to the Ricci scalar.

Summarizing, we have written the full DFT action for the fields $\tilde{g}$, $\beta$ and $\tilde{\phi}$ 
in terms of geometrical quantities that make the invariance under $x$ diffeomorphisms parametrized by 
$\xi^{i}$ manifest. In this formulation the remaining gauge invariance, that is, under the $\tilde{x}$ diffeomorphisms 
parametrized by $\tilde{\xi}_{i}$, is hidden. However, we may now choose different field variables, including 
the original fields $g$ and $b$, and use the `dual' of the geometrical objects introduced above in order 
to make the  $\tilde{\xi}_{i}$ gauge invariance manifest.  Specifically, in addition to $g$ and $b$ we 
introduce the new dilaton $\phi^{\prime\prime}$ according to 
 \be
   \sqrt{|\det{g^{ij}}|}\,e^{-2\phi^{\prime\prime}} \ = \ e^{-2d}\;, 
 \ee 
which in analogy to (\ref{dildef}) is defined such that it transforms as a scalar under  $\tilde{x}$ diffeomorphisms.
We can then define the dual of the full $R$ flux (\ref{Rflux}), which gives a `covariantized $H$ flux' 
 \be
  H_{ijk} \ = \ 3\big(\partial_{[i}b_{jk]}+b_{p[i}\tilde{\partial}^{p}b_{jk]}\big)\;, 
 \ee
and the dual of the anholonomic tilde derivative $\tilde{D}^{i}$, giving 
 \be
  D_{i} \ = \ \partial_{i}-b_{ij}\tilde{\partial}^{j}\;. 
 \ee
Similarly, all other geometrical objects are obtained by systematically sending $\tilde{g}_{ij}\rightarrow g^{ij}$, 
$\beta^{ij}\rightarrow b_{ij}$ and interchanging all upper and lower indices. In particular, 
the conventional Christoffel symbols based on $g_{ij}$ are extended to the dual of the connection 
components (\ref{fullconn}), defining an object $\widecheck{\Gamma}^{k}{}_{ij}$ with which we can  
construct a Ricci scalar as in (\ref{newscalar}). 
The full DFT action can then  be written as         
 \be\label{conjaction3}
 \begin{split}
  S_{\rm DFT} \ = \ \int   dxd\tilde{x}\, \sqrt{|\det{g^{ij}}|}e^{-2\phi^{\prime\prime}}\Big[{\cal R}(g^{ij},\tilde{\partial})&+
  \widecheck{\cal R}(\widecheck{\Gamma}^{k}{}_{ij}) - \frac{1}{12}H_{ijk} H^{ijk} \\ 
  &+4 (\tilde{\partial} \phi^{\prime\prime})^2
  + 4\big(D_i\phi^{\prime\prime}+{\cal T}_i\big)^2 \Big]\;. 
 \end{split}
 \ee
The proof for this expression of the DFT action proceeds in exactly the same way as in section \ref{sec:redef}, 
just with all upper and lower indices interchanged. In this form the invariance under  $\tilde{\xi}_{i}$ gauge 
transformations is manifest. In total, given the two actions  (\ref{conjaction2}) and  (\ref{conjaction3}) 
that make either the $\xi^{i}$ or $\tilde{\xi}_{i}$ gauge invariance manifest, this provides an alternative 
proof for the full gauge invariance of DFT.

\newpage\section{Relating ten- and four-dimensional non-geometric fluxes}\label{sec:4d}

In this section, we first discuss which uplifts of lower-dimensional gauged supergravities our ten-dimensional field redefinition provides. We then turn to the dimensional reduction of the ten-dimensional Lagrangian obtained in this paper, which reproduces the non-geometric terms of the four-dimensional scalar potential. This provides a ten-dimensional origin for these terms, and further motivates why the ten-dimensional $Q$ and $R$ are related to the four-dimensional non-geometric fluxes. We finally come back to the global aspects of this reduction, and the relation to non-geometric field configurations.

\subsection{Gauged supergravity and $O(D,D)$ orbits}\label{sec:gaugedsugra}

String compactifications from ten to lower dimensions in general lead  to supergravity theories, which describe the interactions of
the light modes after integrating out
all massive string excitations.
One important requirement is that the lower-dimensional supergravity action is consistent with the duality symmetries that
act on a given background space. In case the considered backgrounds are completely symmetric under the duality group, i.e.
the duality transformations act as automorphisms on the moduli space of the massless moduli fields (like e.g. for toroidal compactifications),
the corresponding effective actions must be given in terms of  duality invariant functions (automorphic functions) of the scalar moduli fields
\cite{Ferrara:1989bc,Font:1990gx}.
On the other hand, if the duality group acts as transformations between different, but from the string point of view equivalent backgrounds, e.g. a geometric space
is mapped to a non-geometric background as is true for the duality chain in \eqref{eq:TdualityChain}, differently looking supergravity actions will 
be transformed into each other by the action of the duality group on the scalar fields. However these seemingly different 
effective actions are completely equivalent as low energy theories, and in particular the vacuum structure of the theory
will not change within given  orbits of the duality group. Of course, different backgrounds that are not related by duality transformations will
in general lead to physically inequivalent effective descriptions in lower dimensions.

A convenient way to think about gauged supergravity in dimensions $D<10$ is in terms of the
so-called embedding tensor formalism, see \cite{Samtleben:2008pe} and references therein.  Gauged supergravities can be classified by an embedding tensor
$\Theta$ that lives in a certain representation of the global duality group of the ungauged theory.
It encodes which subgroup of the duality group is promoted to a local symmetry and describes the
mass parameters and coupling constants due to the gauging.  The embedding tensor formulation
formally preserves covariance under the full duality group, because $\Theta$ can be thought of as a covariant
tensor, but any choice of a constant non-vanishing embedding tensor breaks the symmetry down to the subgroup
that leaves $\Theta$ invariant. Even though the original duality group is thus not a proper rigid symmetry
of the gauged theory, any two embedding tensors related by a duality transformation lead to physically
equivalent theories. The reason is that by the duality covariance of the embedding tensor formulation
the simultaneous action of a duality group element $h$ on the fields, generically denoted by $\Phi$, and $\Theta$,
leaves the action invariant,
 \be\label{O66redef}
  S_{\rm gauged}\big[\Phi, \Theta] \ = \  S_{\rm gauged}\big[h(\Phi), h(\Theta)]\;.
 \ee
In other words,  the gauged supergravity obtained by sending $\Theta\rightarrow  \tilde{\Theta}= h(\Theta)$ is
equivalent to the original one because they can be related by a field redefinition $\Phi\rightarrow h(\Phi)$.
Thus, physically inequivalent theories are in one-to-one correspondence with the orbits of
the duality group.  However, the gauged supergravities obtained in, say, 4D upon flux compactifications of the
standard 10D supergravity  (\ref{eq:Lns}) do not fill complete orbits under $O(6,6)$ (which generically is part of
the duality group in 4D), as we discussed in the introduction for the missing $Q$ and $R$ flux terms.
Our method provides a higher-dimensional origin for those gauged supergravities that complete the $O(6,6)$
orbit by giving the appropriate field basis, as indicated by the following diagram.
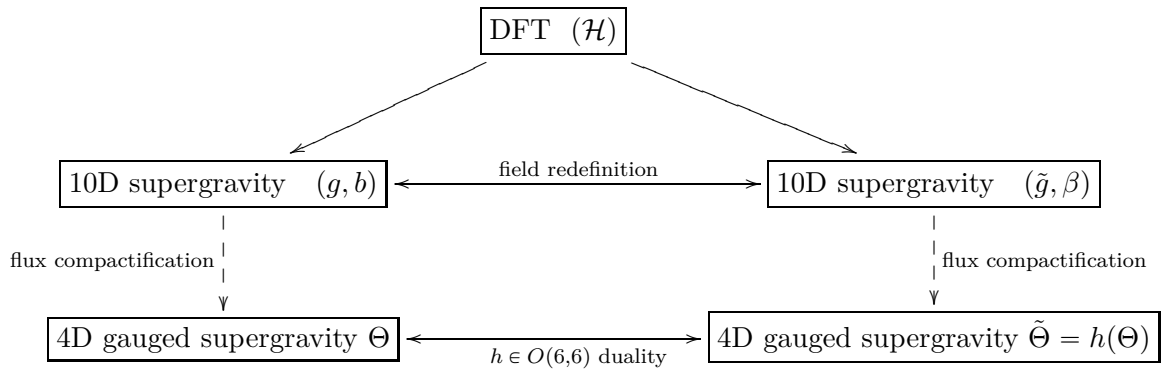
\begin{equation*}
\xymatrix@R=1.2cm{ & \boxed{\text{DFT}~~({\cal{H}})} \ar[dl] \ar[dr] \\
\boxed{\text{10D supergravity }~~(g,b)} \ar@{-->}[d]_{\text{flux compactification}} \ar[rr]^{\text{field redefinition}} & \ar[l] \hspace{2cm} & \boxed{\text{10D supergravity }~~(\tg,\beta)} \ar@{-->}[d]^{\text{flux compactification}} \\
\boxed{\text{4D gauged supergravity } \Theta} \ar[rr]_{h\;\in\; O(6,6) \text{ duality}} & \ar[l] & \boxed{\text{4D gauged supergravity } \tilde{\Theta}=h(\Theta)}}
\end{equation*}
Here, the upper horizontal arrow indicates the field redefinition (\ref{eq:fieldredef}) in 10D that, upon reduction,
corresponds to a field redefinition (\ref{O66redef}). From the point of
view of DFT these different choices of field basis are just different parameterizations of the fundamental
field given by the generalized metric ${\cal H}_{MN}$. In fact, being a symmetric tensor and an $O(D,D)$
group element, ${\cal H}_{MN}$ may be parametrized in terms of symmetric tensor and either a two-form or
an antisymmetric bi-vector, as indicated in (\ref{equality}) and also discussed in \cite{allp11}. While the standard parametrization in terms of
a two-form leads to the usual NSNS action, here we investigate the result of the second parameterization.

According to the above picture we can uplift to 10D those non-geometric fluxes that are T-dual to geometric ones.
However, for a given background, it is to be expected that only one of the 10D descriptions is well-defined, and hence an uplift of the 4D theory requires using this preferred field basis. In order to lift more 4D solutions, and possibly find a complete uplift of the $O(6,6)$ orbit, we need to complement the field redefinition with other T-duality transformations of the background, as in the toroidal example discussed in \cite{allp11}.

It remains an open question whether the new Lagrangian $\tL$, depending on $\tg$ and $\b$, would be of use to uplift other 4D solutions, which are not in an $O(6,6)$ orbit that contains geometric solutions. There might be globally well-defined solutions to the equations of motion derived from $\tL$ that are not related to well-defined solutions corresponding to ${\cal L}$. Concretely, the restriction that a 10D solution is reducible to 4D, via a flux compactification, is a global restriction, and can hence differentiate between two locally equivalent solutions. We come back to this discussion in section \ref{sec:global}.

\subsection{Dimensional reduction}

\subsubsection{Preliminary ideas: the T-duality chain}\label{sec:Tdchain}

For a compactification on a background with isometries, the four-dimensional theory should inherit the T-duality transformation properties from the ten-dimensional theory. Terms in the superpotential of four-dimensional supergravity, as well as the corresponding scalar potential, should transform into each other by T-duality transformations. In this way, it has been realized that new types of terms are needed in the potential, that should be generated by specific discrete quantities ${Q_c}^{ab}$ and $R^{abc}$. Equivalently, one can study gauge algebras of gauged supergravities, in which fluxes enter as structure constants. There as well, these new quantities have been required for the algebra to be T-duality covariant \cite{stw05, dh05}. Note that the different terms of the superpotential transform into each other in a specific order, as summarized by the T-duality chain \eqref{eq:TdualityChain}, that is
\bea
A: H_{abc} \ \
\buildrel {T_a} \over \longleftrightarrow \ \
B: f^{a}{}_{bc} \ \
\buildrel {T_b} \over \longleftrightarrow \ \
C: {Q_c}^{ab} \ \
\buildrel {T_c} \over \longleftrightarrow \ \
D: R^{abc} \ ,\label{eq:Tdchain}
\eea
where $A, B, C, D$ denote the different backgrounds obtained by performing T-duality.\footnote{The geometric flux $f^{a}{}_{bc}$ is given in terms of the vielbein ${e^a}_m$ and its inverse ${e^m}_a$ as
\beq
f^{a}{}_{bc}={e^a}_m \left({e^k}_b\ \del_k {e^m}_c- {e^k}_c\ \del_k {e^m}_b \right) = - 2 {e^k}_{[b} {e^m}_{c]} \del_k {e^a}_m \ .\label{struct}
\eeq
Its appearance in \eqref{eq:Tdchain} rather denotes the potential term obtained by dimensional reduction from the internal Ricci scalar.} This four-dimensional chain \eqref{eq:Tdchain} was first inspired by the famous toroidal example worked out in \cite{kstt02, lnr03} and recalled in \cite{allp11}. In that example, while the field configurations are geometric in situations $A$ and $B$, the second T-duality leads to a non-geometric configuration in the ten-dimensional sense of \cite{hmw02, fww04}. Indeed, the metric and $b$-field in $C$ need a stringy symmetry (in practice a T-duality) to glue when going from one patch to the other. From the four-dimensional perspective, this corresponds to the situation with $Q$. For this reason, $Q$ (and $R$) were named the non-geometric fluxes.

The $R$ flux remains more mysterious in this discussion, because the situation $D$ where it should appear, would be reached in the toroidal example by performing a T-duality along a non-isometry direction. This requires to extend the standard definition of the T-duality. To this end, various proposals have appeared in the literature \cite{eg95, h06a}, including the relation to mirror symmetry \cite{syz96}. In DFT, which is inspired from string field theory \cite{kz92, z92}, T-duality transformations along non-isometry directions are naturally realized by virtue of the doubled coordinates. There, not only the fields get transformed by the $O(D,D)$ element, but also the coordinates do. In particular, when the $O(D,D)$ element is a Buscher transformation in a specific direction, this action on the coordinates results in exchanging $x$ and $\tx$ along this direction. This is obviously not seen when T-dualizing along an isometry direction, since the fields do not depend on the associated coordinate. On the contrary, it has a non-standard effect when T-dualizing along non-isometries. In particular, this allowed in \cite{dpst07, abmn11} to perform a last T-duality in the toroidal example, realizing in ten dimensions the last step of the chain \eqref{eq:Tdchain}.

\subsubsection{Dimensional reduction, scalar potential, and non-geometric terms}

We now want to dimensionally reduce the ten-dimensional Lagrangian \eqref{eq:Lfinal} to four dimensions, to see specifically how the four-dimensional scalar potential emerges. In particular, we would like to verify that our ten-dimensional $Q$ and $R$ give rise precisely to non-geometric potential terms.

We start with a ten-dimensional space-time and restrict ourselves to the set of fields considered so far: a metric, a dilaton, and $b$ or $\beta$. We then split the space-time into a four-dimensional maximally symmetric space-time times a six-dimensional internal space. We consider a compactification ansatz for our fields accordingly. First, the metric is factorized to fit with the product structure, and the four-dimensional metric only depends on the four-dimensional coordinates. Second, any flux constructed out of $b$ or $\beta$ is restricted to have purely internal components. This implies in particular that $b$ and $\beta$ are purely internal, and that they only depend on internal coordinates.

Given this ansatz, we consider only two scalar fields in this dimensional reduction: the volume scalar $\rho$ and the four-dimensional dilaton $\sigma$. This simplified set-up is enough for our purposes, because the appearance of these two fields in the scalar potential is model-independent, and is sufficient to distinguish the non-geometric terms from the others. These fields are defined as follows with respect to the internal metric $g_6$ and the dilaton $\phi$
\beq
g_{6 ij} = \rho\ g_{6 ij}^{(0)}  \ , \quad e^{-\phi}= e^{-\phi^{(0)}} e^{-\varphi} \ , \ \sigma = \rho^{\frac{3}{2}}\ e^{-\varphi} \ , \label{eq:mod}
\eeq
where the index $^{(0)}$ denotes the vacuum expectation values. We then restrict the background metric $g_6^{(0)}$ to depend purely on internal coordinates, and the background dilaton to be constant in order to define the string coupling constant $e^{\phi^{(0)}}=g_s$, while $\rho$ and $\sigma$ depend only on four-dimensional coordinates. Note that the vacuum value of these scalars is obviously $1$. Finally, the other ten-dimensional fields are simply set to their vacuum value.

To illustrate the dimensional reduction, we first describe it for the NSNS action $\frac{1}{2\kappa^2} \int \d^{10} x \ \L$, where the Lagrangian $\L$ is given in \eqref{eq:Lns}. We will then apply it to the ten-dimensional Lagrangian of interest \eqref{eq:Lfinal}. To obtain the reduced four-dimensional theory, we should first insert the fields just defined into the ten-dimensional action. In particular, using \eqref{eq:mod}, the dependence on $\rho$ is easy to determine: it is simply given by the scaling with respect to the internal metric. One gets for instance
\beq
\sqrt{|g_6|} = \rho^3\ \sqrt{|g_6^{(0)}|} \ , \ \R_6=\rho^{-1}\ \R_6^{(0)} \ , \ H_{ijk} H^{ijk} =\rho^{-3}\ H^{(0)}_{ijk} H^{(0)ijk} \ ,
\eeq
where we denote by $\R_6$ the internal Ricci scalar. We define the background volume as $v_0=\int \d^6 x \sqrt{|g_6^{(0)}|}$ and the four-dimensional Planck mass $M_4$ by $M_4^2=v_0/(2 \kappa^2 g_s^2)$. We go to the four-dimensional Einstein frame by scaling the metric $g_4$ with the dilaton $\sigma$ as $g_{4\mu\nu}= \sigma^{-2} g_{\mu\nu}^E$. Eventually, the ten-dimensional NSNS action reduces to the following four-dimensional action in Einstein frame
\beq
S_E= M_4^2 \int \d^4 x \ \sqrt{|g^E|} \left(\R^E_4 + {\rm kin} - \frac{1}{M_4^2} V(\rho,\sigma)  \right) \ ,
\eeq
where ``kin'' denotes the scalar kinetic terms that are not needed explicitly here. The scalar potential $V$ is given by
\beq
V(\rho,\sigma) = \sigma^{-2} \left(\rho^{-3}\ V_H^0 + \rho^{-1}\ V_f^0 \right)\ , \label{eq:potNS}
\eeq
where we defined the following quantities
\beq
V_{H}^0=\frac{M_4^2}{v_0} \int \d^6 x\ \sqrt{|g_6^{(0)}|} \ \frac{1}{12} H^{(0)}_{ijk} H^{(0)ijk}  \ , \ V_{f}^0=-\frac{M_4^2}{v_0} \int \d^6 x \ \sqrt{|g_6^{(0)}|}\ \R_6^{(0)} \ . \label{eq:intglobal}
\eeq
Note that there is an implicit assumption made here: the background fields are supposed to be globally well-defined, so that they can be integrated. We will come back to this point in section \ref{sec:global}.

Now we want to perform a similar dimensional reduction starting with the ten-dimensional Lagrangian \eqref{eq:Lfinal}. The latter depends on the fields $\tg$, $\beta$, and $\tp$, so the scalar fields $\rho$ and $\sigma$ are now defined with respect to $\tg$ and $\tp$. The same holds for the background volume $v_0$, the internal Ricci scalar denoted $\tR_6$ and the string coupling constant $g_s$. Let us recall from the compactification ansatz that $\tp = \tp^{(0)} + \varphi$, where $\tp^{(0)}$ is taken as a constant, while $\varphi$ only depends on four-dimensional coordinates. In addition, $\beta$ is restricted to have only internal components. Therefore, any contraction of the type $\beta^{km} \del_m \tp$ vanishes in this dimensional reduction. In  particular, one gets
\beq
-2 \beta^{km} \del_m d = \beta^{km} \del_m \ln \sqrt{|\tg|} = \frac{1}{2}\beta^{km} \tg^{pq}\del_m \tg_{pq} \ .
\eeq
Out of \eqref{eq:Lfinal}, the only contribution to the dilaton kinetic term is therefore the standard one coming from the ten-dimensional $(\del \tp)^2$.

Let us also consider the terms of the ten-dimensional Lagrangian \eqref{eq:Lfinal} involved in the $Q$  and $R$ flux terms of the potential (as given below in \eqref{eq:termssugra}). Using similar arguments from the compactification ansatz, one can verify that in all these terms, indices and derivatives can be restricted to be internal ones. This implies that the dependence on $\rho$ of these terms is obtained only by scaling arguments, since no derivative acts on it. With these simplifications, we obtain after dimensional reduction the following scalar potential
\beq
V(\rho,\sigma) = \sigma^{-2} \left(\rho^{-1}\ V_f^0 + \rho\ V_Q^0 + \rho^{3}\ V_R^0  \right)\ , \label{eq:potsugra}
\eeq
where
\bea
& V_{f}^0=-\frac{M_4^2}{v_0} \int \d^6 x \ \sqrt{|\tg_6^{(0)}|}\ \tR_6^{(0)} \ ,\label{eq:termssugra}\\
& V_{R}^0=\frac{M_4^2}{v_0} \int \d^6 x\ \sqrt{|\tg_6^{(0)}|} \ \frac{1}{12} R^{(0)ijk} R^{(0)}_{ijk} \ ,\nn \\
& V_{Q}^0=-\frac{M_4^2}{v_0} \int \d^6 x\ \sqrt{|\tg_6^{(0)}|} \ \Bigg( -\frac{1}{4}\tilde{g}_{ik} \tilde{g}_{jl}\tilde{g}^{rs} \,Q_{r}{}^{kl}\,Q_{s}{}^{ij} + \frac{1}{2} \tilde{g}_{pq} Q_{k}{}^{lp} Q_{l}{}^{kq} \nn\\
& \qquad \qquad \qquad \qquad \qquad \qquad + \tilde{g}_{jl}\tilde{g}_{pq}\beta^{jm}\left( Q_{k}{}^{lp}\partial_m\tilde{g}^{kq}+\partial_{k}\tilde{g}^{lp}\, Q_m{}^{kq}\right) \nn\\
& \qquad \qquad \qquad \qquad \qquad \qquad -\frac{1}{4}\tilde{g}_{ik}\tilde{g}_{jl}\tilde{g}_{pq} \left( \beta^{pr}\beta^{qs}\partial_r \tilde{g}^{kl} \partial_s\tilde{g}^{ij} -2 \beta^{ir}\beta^{js} \partial_r \tilde{g}^{lp}\partial_s\tilde{g}^{kq} \right) \nn\\
& \qquad \qquad \qquad \qquad \qquad \qquad +\frac{1}{2\sqrt{|\tg|}}\ \tg^{pq} \del_k \tg_{pq}\ \del_m \left(\sqrt{|\tg|} \tg_{ij} \beta^{ik} \beta^{jm} \right) \Bigg) \ ,\nn
\eea
and the indices $^{(0)}$ and $_6$ are understood on all fields in $V_{Q}^0$.

With this last reduction, we see that we obtain two new types of scaling behaviour with respect to $\rho$. These correspond to the non-geometric terms of the scalar potential. Indeed, it was argued in \cite{hktt08} that the most general potential (from the NSNS sector) should be given by
\beq
V(\rho,\sigma) = \sigma^{-2} \left(\rho^{-3}\ V_H^0 + \rho^{-1}\ V_f^0 + \rho\ V_Q^0 + \rho^{3}\ V_R^0  \right)\ , \label{eq:pot}
\eeq
where $V_Q^0$ and $V_R^0$ are constants depending on the four-dimensional $Q$ and $R$ fluxes, respectively. These two non-geometric terms make this scalar potential T-duality covariant, as discussed in section \ref{sec:Tdchain}. From our dimensional reduction, we obtained expressions for $V_Q^0$ and $V_R^0$ in terms of ten-dimensional fields, as in \eqref{eq:intglobal}. We conclude that the field redefinition performed on the NSNS Lagrangian provides a lift to ten dimensions of the four-dimensional $Q$ and $R$ fluxes, since the corresponding potential terms are reproduced.

Note that for a given background, only some of the four terms of \eqref{eq:pot} could be turned on. For the toroidal example mentioned above, only one out of the four is present. In particular, for the situation $C$, one gets $V_Q^0$ non-zero, and given by the formula \eqref{eq:termssugra}. This was actually already shown in \cite{allp11}. Indeed, the simplifying assumption used in that paper is automatically satisfied in the toroidal example, and our $V_{Q}^0$ then reduces to its first term, the square of the $Q$ flux, while the $R$ flux vanishes. Given the more complete formulas \eqref{eq:termssugra} derived here, it would also be interesting to have an example with a non-trivial $R$.

Let us now have a closer look at the formulas \eqref{eq:termssugra}. The $R$ flux term in \eqref{eq:termssugra} is given by the square of $R^{ijk}=3 \beta^{p[i}\partial_{p}\beta^{jk]}$. By analogy with the $H$ flux term in \eqref{eq:intglobal}, one can view this $R^{ijk}$ as being the ten-dimensional supergravity $R$ flux and corresponding to its four-dimensional counterpart. On the contrary, the $Q$ flux term is more complicated, and it is difficult to identify directly the four-dimensional $Q$ flux there. For this reason, the ten-dimensional $Q_k{}^{mn}$ should here only be understood as a notation, corresponding to the one of \cite{allp11}. The study of DFT diffeomorphisms nevertheless revealed a structure behind $V_Q^0$. Indeed, the formula \eqref{eq:termssugra} of the $Q$ flux term can be derived from the DFT Lagrangian \eqref{conjaction2} up to total derivatives, following section \ref{sec:sugra} and appendix \ref{ap:calc}. More precisely, this four-dimensional term would be obtained from the Ricci scalar $\widecheck{\cal R}|_{\tilde{\del}=0}$, together with the last square term in \eqref{conjaction2}. Therefore, instead of a single square, the four-dimensional $Q$ flux term should rather be thought of as a sum of squares, as are Ricci scalars with constant connections. This is then analogous to the geometric flux term. We already noticed that both $f$ and $Q$ have a mixed index structure, they are not tensors, and they are related to connections. In addition, the standard Ricci scalar is sometimes expressed as a sum of squares of $f^{a}{}_{bc}$, as, for instance, in the case of twisted tori (solvmanifolds) for which the $f^{a}{}_{bc}$ are constant. The same should hold for $Q$ here.

Obtaining $V_{Q}^0$ from the DFT Lagrangian \eqref{conjaction2} which has the Ricci scalar involves total derivatives. Those could actually contribute non-trivially, as will be discussed in section \ref{sec:global}, and therefore modify the expression we gave for $V_{Q}^0$. Put differently, knowing which total derivative should be discarded, as in \eqref{eq:Lageqgen}, and which should be kept (to form a Ricci scalar for instance), is not clear here. It probably depends on the background considered, as we discuss in the following.

\subsection{Global aspects and preferred field basis}\label{sec:global}

A ten-dimensional Lagrangian, as the one considered above, is a local quantity, whereas an action is sensitive to the global aspects through the integration. In dimensional reductions, as discussed previously, one needs to integrate the background fields over the six-dimensional space to get the four-dimensional potential. However, a non-geometric configuration in ten dimensions usually has global issues (for instance the fields are not single-valued). Therefore, it is not clear how to perform this integration. Put differently, the geometry is not the standard one, and so the usual compactification procedure, which could produce the desired four-dimensional potential, cannot be applied. This question was discussed at length in \cite{allp11}, and the field redefinition was again proposed as an answer.

To illustrate this idea, let us first come back to the toroidal example mentioned previously. In the situation C, the metric and the $H$  flux  are ill-defined because we face a non-geometric configuration. Therefore, the integrals \eqref{eq:intglobal} do not really make sense. Equivalently, the associated NSNS Lagrangian is globally ill-defined, preventing to consider the NSNS action. However, by performing the field redefinition to $\tg$ and $\beta$, the new Lagrangian obtained turns out to be well-defined. In particular, the new metric $\tg$ is that of a flat torus, so in a sense, the field redefinition restores the standard notion of geometry. The other quantities in the Lagrangian, as the $Q$ flux for instance, are also well-defined. One can then perform the integration without trouble. This argument is example-based, but as we proposed in \cite{allp11}, it may work as well for other examples, at the possible cost of considering another field redefinition. In general, the situation would be the following:
\beq
\L_{{\rm NSNS}}= \L_{{\rm new}} + \del(\dots) \ . \label{eq:Lageqgen}
\eeq
The NSNS Lagrangian can be rewritten as a new Lagrangian up to a total derivative, where this $\L_{{\rm new}}$ is expressed in terms of the redefined fields (in this paper this is exemplified by $\L_{{\rm new}}$ being \eqref{eq:Lfinal} and the total derivative given by \eqref{totder}). The idea is then that for a non-geometric configuration, while $\L_{{\rm NSNS}}$ would be ill-defined, the field redefinition would be such that $\L_{{\rm new}}$ is well-defined. One can then perform the dimensional reduction with this last Lagrangian, since the associated action and integrals will make sense. This way, the $Q$ flux term of \eqref{eq:pot} was reproduced in \cite{allp11} for the toroidal example.

The field basis which allows to get a well-defined $\L_{{\rm new}}$ in \eqref{eq:Lageqgen} has been named the ``preferred field basis''. The proposal of \cite{allp11} was that this set of fields and associated $\L_{{\rm new}}$ is the proper low energy effective description of string theory on a given non-geometric background. Note that naively, the low energy description is given by the $\L_{{\rm NSNS}}$, but since the two differ by a total derivative \eqref{eq:Lageqgen}, our proposal is then to discard this total derivative. This last point is less trivial than it seems. Indeed, in a non-geometric configuration, $\L_{{\rm NSNS}}$ can be non-single valued. If we have at hand a preferred basis, then $\L_{{\rm new}}$ is single-valued. Therefore, given \eqref{eq:Lageqgen}, the content of the total derivative is not single-valued. It means that it does not integrate to zero. Throwing away the total derivative is then not a trivial statement. This is why there is a difference between choosing one field basis or the other, and their associated actions, as the low energy effective description of string theory. Thus we propose that there should be a preference according to whether the background is geometric or not. To make such a proposal more concrete, a world sheet perspective on this question could be useful \cite{h06b, bhm07, h09}.

\section{Conclusions and Outlook}\label{sec:conc}

Since the introduction of the non-geometric $Q$ and $R$ fluxes in four-dimensional gauged supergravity, a wealth of studies has been devoted to better understand their properties. The subsequent realization that these fluxes help in the constructions of phenomenologically interesting four-dimensional solutions has further fuelled the interest in this subject. One of the main questions raised by these studies is whether these four-dimensional solutions have a higher-dimensional description in string theory.

In this paper we have taken a first step towards such a realization. By studying the NSNS action, and its generalization to strongly constrained DFT, we have shown how a change of field basis, replacing the NSNS fields $g_{ij}, b_{ij},\phi$ with a new metric $\tg_{ij}$, an antisymmetric bi-vector $\b^{ij}$ and a new dilaton $\tp$, gives rise to a new action
which contains $Q$ and $R$. The field redefinition can equally well be applied to ten-dimensional supergravity or to DFT, as we have shown by performing both analyses. However, since the change of field basis takes the form of a T-duality $O(D,D)$ transformation in all directions, the computations are greatly simplified using DFT, which can be formulated in an $O(D,D)$ invariant fashion. 

Using the DFT framework, we can also give a precise geometrical meaning to the non-geometric fluxes. Concretely, the DFT action for the new fields takes the form
\be \label{eq:sdftfinal}
  S_{\rm DFT} \ = \ \int   dxd\tilde{x}\,\sqrt{|\tg|}\,e^{-2\tilde{\phi}}\Big[{\cal R} + \widecheck{\cal R}-\frac{1}{12}R^{ijk} R_{ijk} 
  +4 (\partial \tilde{\phi})^2
  + 4\big(\tilde{D}^i\tilde{\phi}+{\cal T}^i\big)^2  \Big]
   \; ,
 \ee 
where the standard Ricci scalar ${\cal R}$ and dilaton kinetic term $(\partial \tp)^2$ are accompanied by a dual Ricci scalar $\widecheck{\cal R}$, an $R$ flux term, and a dual kinetic term for the dilaton $\tilde{\phi}$ that includes the new torsion ${\cal T}^i$.  All terms in this action are separately covariant under  the DFT diffeomorphisms of the $x$ coordinates, and hence the corresponding quantities have a clear geometrical meaning. Particularly,  we have identified the $R$ flux with the covariant field strength of $\b$, and so the $R^2 $-term is manifestly covariant. The geometric interpretation of $Q$ is more subtle; we find that it is the antisymmetric part of a dual connection, and therefore appears in the action as a part of the dual Ricci scalar $\widecheck{\cal R}$ and the torsion ${\cal T}^i$. Thus, the $R$ flux is a three-form with respect to the dual coordinates $\tx$ and $Q$ is related to a connection. This should not come as a surprise,  since it precisely matches the dual situation for the geometrical fluxes, where $H$ is a three-form and $f$ is related to the Levi--Civita spin connection.

We have checked, by performing a model-independent dimensional reduction, that the higher-dimensional $Q$ and $R$ flux produce the expected non-geometric flux terms in the four-dimensional scalar potential. This check further strengthens our higher-dimensional identification of the flux terms, and shows that we have found ten-dimensional lifts of four-dimensional gauged supergravity solutions that were previously lacking. We have restricted our analysis to supergravity or strongly constrained DFT, whereas some lower-dimensional gravities have been shown to correspond to situations where the strong constraint is relaxed \cite{Dibitetto:2012rk} (see also \cite{abmn11,g11,gm12}). Consequently, our lift is not exhaustive. We believe, however, that the structures found here, and in particular the geometric interpretation of the $Q$ and $R$ fluxes, will be a guide for more general treatments of non-geometric situations.

In this paper, we have focused on the NSNS sector of DFT and supergravity. In order to study explicit compactifications and derive four-dimensional solutions, further ingredients must be added to the theory. In heterotic compactifications, gauge fields should be considered. In geometric type II solutions, Ramond-Ramond (RR) fluxes as well as $D$-branes and orientifold planes play an important role, and the same is expected for non-geometric solutions. It would be interesting to include these degrees of freedom into our analysis, for example along the lines of \cite{Hohm:2011zr,Hohm:2011dv,Coimbra:2011nw}, and thus complete our final action \eqref{eq:sdftfinal}. Moreover, using such a completed action, or its supergravity version, we could solve the corresponding ten-dimensional equations of motion and look for concrete compactifications. One interesting set of examples to study would then be generalized Calabi--Yau compactifications with non-geometric fluxes. For elliptically fibered Calabi--Yau manifolds, mirror symmetry reduces to T-duality \cite{syz96}, and it is possible that the field redefinition we propose is useful for the study of their non-geometric duals \cite{glmw02, fmt03, gs07}. 

\medskip
To conclude our discussion, let us comment on the possible relation between the geometrical formulation of the  non-geometric $Q$ and $R$ fluxes to the non-commutative and non-associative structures, which are present in non-geometric string backgrounds. Non-commutative and non-associative algebras have been found both for open strings ending on D-branes 
\cite{Chu:1998qz, Schomerus:1999ug, Seiberg:1999vs,Ardalan:1998ks,Cornalba:2001sm,gs06,Ellwood:2006my}
as well for closed strings moving on non-geometric backgrounds \cite{bp11,l10,bdlpr11,Blumenhagen:2011yv,Condeescu:2012sp}.
In an example where a two-torus is fibered over a base circle, it has been shown that a non-commutative algebra for the string coordinates of the fibre torus emerges if there is a  $Q$ flux present, whereas the algebra for all three coordinates becomes non-associative in the presence of an $R$ flux.\footnote{These non-commutative and non-associative structures also appeared in the more mathematically oriented literature  \cite{Bouwknegt:2000qt,Bouwknegt:2004ap,Mathai:2004qq,Mathai:2004qc,Brodzki:2007hg,Saemann:2012ex}, where twisted K-theory is applied to characterize non-geometric backgrounds with D-branes and $B$-fields.}

Let us therefore try to relate the geometrical objects $\b^{mn}$, $Q_m{}^{nl}$ and $R^{mnl}$ discussed in this paper to the deformation parameters of the associated non-commutative respective non-associative algebras.
One first relevant observation in this context is that these objects are closely related to the non-commutative open string geometry on these spaces. In fact, by comparing with eq.~(2.5) in  \cite{Seiberg:1999vs}, one can easily convince oneself that the dual metric $\tilde g$ and the bi-vector $\b$ correspond, respectively, to the open string metric and the open string non-commutativity deformation parameter defined in this reference (see also \cite{h08}).  Concretely, for D2-branes that are wrapped around the torus fibre of the $Q$ flux space,
one obtains the following equal-time commutator for the open string coordinates (at the location $\sigma = 0,\pi$ of the D2-brane):
\begin{equation}\label{commopen}  [ X^m(\tau), X^n(\tau) ]_{\rm open}=     \b^{mn} \, . 
\end{equation}
This defines a so-called Poisson structure in analogy to the momentum algebra of a point particle
moving in a (constant) magnetic field.

Now let us turn to the non-commutative geometry of closed strings moving in the non-geometric $Q$ flux background.
A first guess could be that the non-commutativity is again directly related to the bi-vector $\b$, leading to the same algebra
(\ref{commopen}) as for the open strings.
However, this will not be quite correct: as discussed in \cite{l10,Condeescu:2012sp}, only an extended closed string which is wrapped $\tilde p^k$ times around the base of the fibration is sensitive to the global ill-definedness of the two-dimensional fibre torus. As a result the fibre geometry becomes non-commutative with non-commutativity deformation parameter given in terms of  the winding number $\tilde p^k$: 
\begin{equation}\label{commclosed} 
 [ X^m(\tau,\sigma), X^n(\tau,\sigma) ]_{\rm closed}\sim   \epsilon_k{}^{mn}\tilde p^k \, .
\end{equation}
In view of this result, we propose the following integral relation between the non-geometric $Q$ flux and the closed string non-commutativity:
\begin{equation}\label{commcloseda}
 [ X^m(\tau,\sigma), X^n(\tau,\sigma) ]_{\rm closed}=  \oint_{C_k}Q_k{}^{mn}(X)~ dX^k \ ,
\end{equation}
where $C_k$ is a non-trivial homology base cycle, around which the closed string is wrapped $\tilde p^k$ times. In the case of constant flux $Q_k{}^{mn}=Q\epsilon_k{}^{mn}$ and $C_k=S^1$ one gets
\begin{eqnarray}\label{commclosedb}
 [ X^m(\tau,\sigma), X^n(\tau,\sigma) ]_{\rm closed}= \oint_{C_k}Q_k{}^{mn} dX^k \;
 = \; 2\pi Q  \epsilon_k{}^{mn}\tilde p^k    \, ,
\end{eqnarray}
in agreement with (\ref{commclosed}). We hope to come back to the relation between $Q$ flux and non-commutativity in future work.

Finally let us discuss the $R$ flux background obtained by a T-duality transformation, $X^k\leftrightarrow \tilde X_k$, in the $k^{\rm th}$ direction from the previous case.
The corresponding closed string background becomes non-associative, as discussed in \cite{bp11} in the context of the $SU(2)$
Wess--Zumino--Witten model, and investigated  in 
\cite{bdlpr11} by  the computation of conformal field theory  amplitudes in the chain of T-dual $H,f,Q,R$-backgrounds
leading to a non-associative algebra of closed string vertex operators.
So it is quite natural to conjecture that the non-geometric flux  $R$  corresponds to the parameter that
controls 
the violation of the Jacobi identity, i.e. to the deformation parameter of the non-associative algebra of the $R$ flux backgrounds:
\begin{equation}\label{nonass} 
 [[ X^m(\tau,\sigma), X^n(\tau,\sigma)],X^k(\tau,\sigma)]_{\rm closed} +{\rm perm.} =     R^{mnk}\ .
 \end{equation}
Note that this non-associativity relation can be at least formally derived from the commutator (\ref{commclosedb}) by using the Heisenberg commutation relation
$[X^k,p^k]=i$
in the $k^{\rm th}$ direction
\cite{l10}.\footnote{Eqs.(\ref{commclosedb}) and (\ref{nonass}) together with $[X^k,p^k]=i$ now define a so-called twisted Poisson structure. 
This structure also emerges
for the momenta of  point particles moving in the field of a magnetic monopole, as it was originally remarked in 
\cite{Jackiw:1984rd} and further discussed e.g. in \cite{Klimcik:2001vg,Alekseev:2004np,Kotov:2004wz};
the magnetic $B$-field of the monopole corresponds to the bi-vector $\b$ in our case,
and the non-closure of $B$ corresponds to the $R$ flux.}

It would be interesting to expand on these connections between  non-geometric fluxes and non-commutative and non-associative string backgrounds in future work. We believe that our higher-dimensional expressions for the $Q$ and $R$ fluxes, which have a clear geometrical meaning, will be helpful in this regard.

\section*{Acknowledgments}
\vspace{-0.25cm}
For useful comments and discussions we would like to thank I.~Bakas, R.~Blumenhagen, D.~Roest, D.~Waldram, and B.~Zwiebach. 
This work is supported by the Alexander-von-Humboldt foundation, the
DFG Transregional Collaborative Research Centre TRR 33
and the DFG cluster of excellence `Origin and Structure of the Universe'.
DL thanks the Simons Center for Geometry and Physics for hospitality. OH and ML thank the Isaac 
Newton Institute in Cambridge, as part of the programme on the Mathematics and Applications of Branes in String and M-theory,  for hospitality.

\newpage
\begin{appendix}
\section{Computational details}\label{ap:calc}

This appendix contains some of the computational details of section \ref{sec:DFTrewrite} and \ref{sec:nongeoflux}. Specifically, we work out the various structures in the DFT Lagrangian \eqref{eq:Ldfttilde} in terms of $\tg, \beta$ and $d$ and the derivative $\tilde{D}^{i}$ \eqref{eq:Dtilde}. We then show that this Lagrangian corresponds to the geometric action \eqref{conjaction}.

Using \eqref{eq:calD2}, the quadratic dilaton term in the DFT action becomes 
\be
\label{eq:b1}
 4\tilde{g}_{ij}\tilde{\cal D}^{i}d\,\tilde{\cal D}^{j}d \ = \ 4\big(\tilde{g}^{ij}\partial_id\partial_jd+\tilde{g}_{ij}
 \tilde{D}^{i}d\tilde{D}^{j}d\big)\;,
\ee  
where the cross-terms $\partial_j d {\tilde{D}}^{j} d$ vanish because of the constraint \eqref{eq:Dconstraint}. The off-diagonal dilaton terms give rise to
\be
\label{eq:b2}
  \begin{split}
   \tilde{g}_{ik}\tilde{g}_{jl}\tilde{\cal D}^{i}d\,\tilde{\overline{\cal D}}{}^{j}\tilde{\cal E}^{kl}
   + \tilde{g}_{ik}\tilde{g}_{jl}\,\tilde{\overline{\cal D}}{}^{i}d\,\tilde{\cal D}^{j}\tilde{\cal E}^{lk} \ = \ 
   -2\big[& \partial_k d\,\partial_l \tilde{g}^{kl}+\tilde{g}_{jl}\,\partial_k d\, \tilde{D}^{j}\beta^{kl}  \\
   &-\tilde{g}_{ik}\,\tilde{D}^{i}d\,\partial_l\beta^{kl}-\tilde{g}_{ik}\tilde{g}_{jl}\tilde{D}^{i}d\,\tilde{D}^{j}\tilde{g}^{kl} \big]\;.
  \end{split}
\ee  
The first term quadratic in $\tilde{\cal E}$ reduces to
  \be
  \label{eq:b3}
  \begin{split}
   -\frac{1}{4} \,\tilde{g}_{ik}\,\tilde{g}_{jl}  \,\tilde{g}_{pq} \, 
     \tilde{{\cal D}}^{p}\tilde{{\cal E}}^{kl}\,  \tilde{{\cal D}}^{q}\tilde{{\cal E}}^{ij} \ = \ &-\frac{1}{4} \,\tilde{g}_{ik}\,\tilde{g}_{jl} \tilde{g}^{rs}\partial_r\tilde{g}^{kl}\,\partial_s\tilde{g}^{ij}
     -\frac{1}{4}\,\tilde{g}_{ik}\,\tilde{g}_{jl} \tilde{g}^{rs} \partial_{r}{\beta}^{kl}\,\partial_{s}{\beta}^{ij}\\
     &- \frac{1}{4}\,\tilde{g}_{ik}\,\tilde{g}_{jl}\tilde{g}_{pq}\big(\tilde{D}^{p}\tilde{g}^{kl}\,\tilde{D}^{q}\tilde{g}^{ij}+\tilde{D}^{p}\beta^{kl}\,\tilde{D}^{q}\beta^{ij}\big)\;, 
  \end{split}
  \ee
where the strong constraint \eqref{eq:strong} was used to cancel some terms. Finally, for the sum of the last two structures in \eqref{eq:Ldfttilde} we get 
\be
\label{eq:b4}
\begin{split}
 \frac{1}{4}\tilde{g}_{ik}\tilde{g}_{jl}\tilde{g}_{pq}\big(\tilde{\cal D}^{i}\tilde{\cal E}^{lp}\,\tilde{\cal D}^{j}\tilde{\cal E}^{kq}
 +\tilde{\overline{\cal D}}{}^{i}\tilde{\cal E}^{pl}\,\tilde{\overline{\cal D}}{}^{j}\tilde{\cal E}^{qk}\big)  \ = \ 
 &+ \frac{1}{2}\tilde{g}_{pq}\,\partial_k\tg^{lp}\,\partial_l\tg^{kq}
 + \frac{1}{2}\tilde{g}_{pq}\,\partial_k\beta^{lp}\,\partial_l\beta^{kq}\\
  &+\frac{1}{2}\tilde{g}_{ik}\tilde{g}_{jl}\tilde{g}_{pq}\big(\tilde{D}^{i}\tilde{g}^{lp}\tilde{D}^{j}\tilde{g}^{kq}
 +\tilde{D}^{i}\beta^{lp}\tilde{D}^{j}\beta^{kq}\big)\\
 &-\tilde{g}_{jl}\tilde{g}_{pq}\big(\partial_k\beta^{lp}\,\tilde{D}^{j}\tilde{g}^{kq}+\partial_{k}\tilde{g}^{lp}\tilde{D}^{j}\beta^{kq}\big)\;. 
\end{split}
\ee

The four structures \eqref{eq:b1}--\eqref{eq:b4} contain some terms that are independent of $\beta$ and only contain standard derivatives. These are exactly the terms we would get if we would set $\tilde{\partial}=0$ and $b=0$ in the original DFT action \eqref{eq:sdft}, but with the fields replaced by their tilded counterparts. Consequently, they combine to the Ricci scalar and the standard kinetic term for the dilaton, up to a total derivative (see \cite{hhz10a} for details on this computation)
\be
\label{eq:b5}
\begin{split}
&4\tilde{g}^{ij}\partial_id\partial_jd
-2 \partial_k d\,\partial_l \tilde{g}^{kl}
-\frac{1}{4} \,\tilde{g}_{ik}\,\tilde{g}_{jl} \tilde{g}^{rs}\partial_r\tilde{g}^{kl}\,\partial_s\tilde{g}^{ij}
+ \frac{1}{2}\tilde{g}_{pq}\,\partial_k\tg^{lp}\,\partial_l\tg^{kq} \\
&= {\cal R}(\tg) 
+4 (\partial \tp)^2    
- \partial_k \left( e^{-2d} \left( -\partial_l \tilde{g}^{lk} - \tilde{g}^{ij} \tilde{g}^{lk} \partial_l \tilde{g}_{ij} \right) \right)
\; .
\end{split}
\ee
As result, we find that the DFT Lagrangian is
\bea
e^{2d} \LD (\tg, \beta, d) &= {\cal R}
+4(\partial \tp)^2
+4(\tilde{D} d)^2    
-\frac{1}{4}\tilde{g}_{ik}\tilde{g}_{jl}\tilde{g}_{pq}\big(\tilde{D}^{p}\beta^{kl} \tilde{D}^{q}\beta^{ij}
   -2\tilde{D}^{i}\beta^{lp}\tilde{D}^{j}\beta^{kq}\big) \nn \\
   &+2\tilde{g}_{ik}\tilde{g}_{jl}\tilde{D}^{i}d \tilde{D}^{j}\tilde{g}^{kl}-2\tilde{g}_{jl}\partial_kd\, \tilde{D}^{j}\beta^{kl}
   +2\tilde{g}_{ik}\tilde{D}^{i}d\,\partial_{l}\beta^{kl} \nn\\
   &-\frac{1}{4}\tilde{g}_{ik}\tilde{g}_{jl}\tilde{g}^{rs}\,\partial_{r}{\beta}^{kl}\,\partial_{s}{\beta}^{ij}+\frac{1}{2}\tilde{g}_{pq}\partial_{k}{\beta}^{lp}\partial_{l}{\beta}^{kq} \nn \\
   &-\tilde{g}_{jl}\tilde{g}_{pq}\big(\partial_{k}{\beta}^{lp}\tilde{D}^{j}\tilde{g}^{kq}+\partial_{k}\tilde{g}^{lp}\, \tilde{D}^{j}\beta^{kq}\big) \label{eq:Ldftcompare}
\\
   &-\frac{1}{4}\tilde{g}_{ik}\tilde{g}_{jl}\tilde{g}_{pq}\big(\tilde{D}^{p}\tilde{g}^{kl} \tilde{D}^{q}\tilde{g}^{ij}
   -2\tilde{D}^{i}\tilde{g}^{lp}\tilde{D}^{j}\tilde{g}^{kq}\big) \nn \\
   &+ e^{2d}\ \partial_k \left( e^{-2d} \left( \partial_l \tilde{g}^{lk} + \tilde{g}^{ij} \tilde{g}^{lk} \partial_l \tilde{g}_{ij} \right) \right)\; . \nn
\eea
For later convenience we integrate the second row of this expression by parts, thus removing the terms that are linear in dilaton derivatives. This results in the Lagrangian \eqref{eq:Ldfttilde1} in section \ref{sec:DFTrewrite}. 

It is now straightforward to identify the above Lagrangian with the geometric DFT action \eqref{conjaction}.  Starting with the Lagrangian  \eqref{eq:Ldfttilde2}, we expand the $(\tilde{D} d)^2$ term in $\tp$ and $\tg$, and integrate the result by parts. After some simplifications we find 
 \bea
e^{2d}  \LD (\tg, \beta, d) &=  {\cal R} 
+4(\partial \tp)^2
+4(\tilde{D} \tp)^2  
-\frac{1}{12}R^{ijk}R_{ijk} \nn \\      
   &-\frac{1}{4}\tilde{g}_{ik}\tilde{g}_{jl}\tilde{g}^{rs}\,Q_{r}{}^{kl}\,Q_{s}{}^{ij}
   -\frac{1}{2}\tilde{g}_{pq}Q_{k}{}^{lp}Q_{l}{}^{kq}
   -\tilde{g}_{ij}\,Q_{p}{}^{pi}\,Q_{q}{}^{qj} \nn \\
   &-2 Q_{l}{}^{lk}  \tilde{D}^{i}\tilde{g}_{ik}
   -\tilde{g}_{jl}\tilde{g}_{pq}\, Q_{k}{}^{lp}\tilde{D}^{j}\tilde{g}^{kq} 
   + Q_p{}^{ip}\tg_{ij}\tg^{kl} \tilde{D}^{i}\tg_{kl} \\
    &-\tilde{D}^i\tilde{D}^j\tilde{g}_{ij}-\tilde{D}^i \left[\tg_{ij}\tg^{kl} \tilde{D}^{i}\tg_{kl}\right]
 -2\tilde{g}_{jl}\,\tilde{D}^{j}Q_{k}{}^{kl}\nn \\
    &-\frac{1}{4}\tilde{g}_{ik}\tilde{g}_{jl}\tilde{g}_{pq}\big(\tilde{D}^{p}\tilde{g}^{kl} \tilde{D}^{q}\tilde{g}^{ij}
   -2\tilde{D}^{i}\tilde{g}^{lp}\tilde{D}^{j}\tilde{g}^{kq}\big) 
   -\frac{1}{4} \tg_{ij}\tg_{kl}\tg_{mn} \tilde{D}^{i}\tg^{kl} \tilde{D}^{j}\tg^{mn}\nn \\
  &+e^{2d}\ \tilde{\partial}^i \left[e^{-2d} \left(\tilde{D}^{j} \tg_{ij}- \tg_{ij} Q_l{}^{jl} + \tg_{ij}\tg^{kl} \tilde{D}^{i}\tg_{kl} \right) \right]  \nn \\
  &+e^{2d}\ \partial_k \left[e^{-2d} \left( \tg_{jl}\tilde{D}^j \beta^{kl} - \beta^{ik} (\tilde{D}^j \tg_{ij} - \tg_{ij} Q_l{}^{jl}) \right.\right. \nn\\
  & \quad\quad\quad\quad\quad \left.\left.+\partial_l \tilde{g}^{lk} + \tilde{g}^{ij} \tilde{g}^{lk} \partial_l \tilde{g}_{ij} 
  -  \beta^{ik} \tg_{ij}\tg^{ml} \tilde{D}^{i}\tg_{ml} \right) \right] 
 \; .\nn
 \eea  
Using \eqref{finalRicci} and \eqref{newscalar} we find 
\bea
\widecheck{\cal{R}} & =
-\frac{1}{4} \tg_{ij} \tg_{mn} \tg^{kl} Q_k{}^{mi}Q_l{}^{nj}
-\frac{1}{2}\tg_{ij}Q_k{}^{lj}Q_l{}^{ki}
-\tg_{ij}Q_k{}^{ki}Q_l{}^{lj} \nn \\
&+ 2 Q_l{}^{lk} \tilde{D}^i \tg_{ik} 
- \tg_{jl} \tg_{pq} Q_k{}^{lp}  \tilde{D}^j \tg^{kq}
+ Q_p{}^{jp} \tg_{ij} \tg^{kl} \tilde{D}^i \tg_{kl}  \nn \\
&-\tilde{D}^i\tilde{D}^j \tg_{ij} + \tilde{D}^i \left[\tg_{ij} \tg^{kl} \tilde{D}^j \tg_{kl} \right] + 2 \tg_{ij} \tilde{D}^i Q_p{}^{pj} \nn \\
&+\frac{1}{4}\tg_{ij} \left(
\tilde{D}^i \tg_{kl} \tilde{D}^j \tg^{kl} - 2 \tilde{D}^i \tg_{kl} \tilde{D}^k \tg^{lj} - \tg_{kl} \tg_{mn} \tilde{D}^i \tg^{kl} \tilde{D}^j \tg^{mn}
\right)  
 \, ,
\eea 
 and hence we have, up to total derivatives,
  \bea
e^{2d}  \LD (\tg, \beta, \tp) &=  {\cal R} + \widecheck{\cal{R}} 
+4(\partial \tp)^2
+4(\tilde{D} \tp)^2    
-\frac{1}{12}R^{ijk}R_{ijk} \nn \\      
&-4 \tilde{D}^i \left[Q_l{}^{lk} \tg_{ik}\right]
-2 \tilde{D}^i \left[\tg_{ij} \tg^{kl} \tilde{D}^j \tg_{kl} \right]
 \; .
 \eea  
The last two terms of this expression can be rewritten in terms of the new torsion ${\cal T}^{i}$:
\beq
-4 \tilde{D}^i \left[Q_l{}^{lk} \tg_{ik}\right]
-2 \tilde{D}^i \left[\tg_{ij} \tg^{kl} \tilde{D}^j \tg_{kl} \right] = 4 \left(\tilde{\nabla}^{i}{\cal T}_{i}-{\cal T}^{i}{\cal T}_{i} \right) \; .
\eeq
We have thereby shown that the DFT Lagrangian corresponds to the geometric action \eqref{conjaction}.

Let us finally record a relation that is useful when going from \eqref{conjaction2} to the second supergravity action \eqref{tildeaction}:
\beq \label{eq:Dphiexpand}
e^{-\tp} \sqrt{|\tg|}\left(\tilde{D}^i \tp + {\mathcal{T}}^i\right) =  \tilde{\partial}^i \left( e^{-\tp} \sqrt{|\tg|} \right) + \partial_m \left( \beta^{mi} e^{-\tp} \sqrt{|\tg|} \right)  \ .
\eeq

\section{Rewriting of the NSNS Lagrangian}\label{ap:rewrite}

In this appendix, we rewrite, as discussed in the Introduction, the NSNS Lagrangian \eqref{eq:Lns} by simply replacing the NSNS (untilded) fields by their expressions in terms of the new tilded fields. The same method was used in \cite{allp11}, so we make use of some partial results from this paper.

Before starting the computation, let us give a few useful relations. It was detailed in \cite{allp11} how equation \eqref{equality} is equivalent to the relations
\bea
g &= \left(\tg^{-1} - \b \tg \b \right)^{-1} = (\tg^{-1} \pm \b)^{-1} \tg^{-1} (\tg^{-1} \mp \b)^{-1} \ , \  \nn \\
&= \frac{1}{2}\left((\tg^{-1}+ \b)^{-1} + (\tg^{-1}- \b)^{-1} \right) \ ,\label{ghat}\\
b&= -(\tg^{-1} \pm \b)^{-1} \b (\tg^{-1} \mp \b)^{-1} \ ,  \nn \\
&= \frac{1}{2}\left((\tg^{-1}+ \b)^{-1} - (\tg^{-1}- \b)^{-1} \right) \ . \label{Bhat}
\eea
From \eqref{ghat} and \eqref{Bhat}, one can easily get the converse relations defining $\tg$ and $\b$, and in particular \eqref{eq:relation}. As in \cite{allp11}, we introduce for later convenience the notation
\beq
G_{\pm}^{mn}=\tg^{mn}\pm\b^{mn} \ , \ (G^{-1}_{\pm})_{mn}=\left((\tg^{-1}\pm \b)^{-1} \right)_{mn} \ , \label{eq:defG}
\eeq
where one can notice that $G_{\pm}^T=G_{\mp}$. This property will allow us in the following to use mainly $G_+$, that we will denote for simplicity as $G=G_+$.\footnote{With respect to the DFT notation used in the bulk of this paper, $G=\tilde{\cal E}$.} The definition \eqref{eq:defG} allows us to rewrite \eqref{ghat} and \eqref{Bhat} as
\beq
g_{mn}=(G^{-1}_{\pm})_{mk} \tg^{kp} (G^{-1}_{\pm})_{np}\ , \ \ g^{mn}=G_{\pm}^{mk} \tg_{kp} G_{\pm}^{np}\ ,\ \ b_{mn} = -(G^{-1}_{\pm})_{mk} \b^{kp} (G^{-1}_{\mp})_{pn} \ . \label{eq:convrewr}
\eeq
Equivalently, one has $g_{mn}=(G^{-1}_{+})_{km}\tg^{kp}(G^{-1}_{+})_{pn} \ , \ g^{mn}=G_+^{km}\tg_{kp}G_+^{pn}$, using the $\pm$ freedom, and the transpose of $G$. Finally, note that from \eqref{eq:convrewr}, we get $(\det (g))^{-1}= (\det (G))^2 \det(\tg)$, which implies that $\det (g)$ and $\det (\tg)$ have the same sign. This will be useful in the following, in particular for the dilaton definition.

Given a matrix $A$ of coefficient $A^{pq}$, we will also make use of the following formulas
\bea
& \ln (\det (A))={\rm tr} (\ln(A))\ , \label{lndet}\\
& \partial_{m} {\rm tr} (\ln(A))={\rm tr} (A^{-1}\partial_m A)\ , \label{eq:lndet}\\
& A^{pq}\left(\partial_k A^{-1}_{qr}\right)=-A^{-1}_{qr}\left(\partial_k A^{pq}\right)\ ,
\eea
valid for an invertible $A$, independently of its signature (using a complex $\ln$ if needed). By convention, a derivative acts on the first object on its right, unless brackets are used.

We can now use the expressions \eqref{ghat}, \eqref{Bhat} and \eqref{eq:dilintro} to rewrite\footnote{\label{foot:apchangeconv} We change conventions with respect to \cite{allp11} by taking $\b \rightarrow -\b$. As we can see from \eqref{ghat} and \eqref{Bhat}, $g$ is independent of this sign while $b$ gets a global minus sign. The computation of the Christoffel symbols, Ricci scalar, and dilaton terms, all rely on the replacement of $g$ by its expression \eqref{eq:convrewr}. Since this relation is independent of the change of sign, the same goes for these computations. A fortiori, the treatment of the second order derivative terms and the total derivative are also unaffected by the change of convention. Finally, the minus in $b$ leads to a global minus sign of the $H$  flux component, which is of no consequence in $H^2$. Therefore, the whole computation of this appendix is independent of the change of convention, and so is the final result.} the NSNS Lagrangian in terms of the variables $\tg$, $\b$ and $\tp$.

\subsection*{Ricci scalars for $g$ and $\tg$}

We start by recalling the definitions needed here. For a generic metric $g_{mn}$ with Levi-Civita connection, one has for the connection coefficients
\beq
2\G_{mkn}=\left(\del_k g_{mn}+\del_n g_{mk}-\del_m g_{kn}\right) \ ,\ {\G^p}_{kn}=g^{pm} \G_{mkn}={\G^p}_{nk} \ , \ {\G^{pq}}_{n}=g^{qk} {\G^p}_{kn} \ .\label{LCdef}
\eeq
Then the Ricci scalar is given by
\bea
\R(g)& =g^{ln} \del_k{\G^k}_{nl} - g^{lp} \del_p {\G^k}_{kl} + {\G^{pn}}_n {\G^k}_{kp} - {\G^{pn}}_k {\G^k}_{np} \label{Riccidef}\\
&= g^{lm}g^{ku} \del_k \del_m g_{lu} - g^{lu}g^{km} \del_k\del_m g_{lu} \label{tR}\\
& +\frac{1}{2} \del_m g_{ln} \del_k g_{pu} \left( 2 g^{kl} g^{mn} g^{pu} - \frac{1}{2} g^{km} g^{ln} g^{pu} + \frac{3}{2} g^{km} g^{np} g^{lu}
 - g^{mp} g^{kn} g^{lu} - 2 g^{mn} g^{kp} g^{lu} \right) \ .\nn
\eea
Each of the four terms in \eqref{Riccidef} was given explicitly in terms of the metric in \cite{allp11}, and out of them, one gets the resulting expression \eqref{tR} for the Ricci scalar.

Now, we want to compute the NSNS Ricci scalar $\R(g)$ for $g$ given in (\ref{eq:convrewr}). To do so, we compute the four terms of the definition \eqref{Riccidef} in terms of the tilded fields and get
\bea
g^{lm} \del_m \G^{k}{}_{kl}&= -\frac{1}{2} g^{lm} \del_m (\tg^{pq} \del_l \tg_{pq} ) + g^{lm} \iG_{qr} \tg^{rp} \left(\tg^{uq} \del_m \del_l \tg_{pu} - \tg_{pu} \del_m \del_l \b^{uq} \right) \\
& + g^{lm} \iG_{qr} \tg^{rp} \del_m G^{qu} \del_l \tg_{pu} + g^{lm} \iG_{qu} \iG_{rk} \del_l G^{kq}  \left( \del_m G^{ur} + \tg^{us} G^{vr} \del_m \tg_{sv} \right) \ ,\nn
\eea
\bea
\G^{pn}{}_n \G^{k}{}_{kp} &= -  \left(\iG_{ru} \del_p G^{ur} + \frac{1}{2} \tg^{rs} \del_p \tg_{rs} \right) g^{pm}
\left(\iG_{lv} \del_m G^{vl} + \frac{1}{2} \tg^{kn} \del_m \tg_{kn} \right) \\
& -\frac{1}{2} \tg^{ml} \del_p \tg_{ml} \del_k \tg_{rs} \left(\tg^{rk}\tg^{sp} -\b^{sp}\b^{rk} \right) + \frac{1}{2} \tg^{ml} \del_p \tg_{ml} \left(\tg_{rs} \b^{sk} \del_k \b^{rp} + \tg_{rs} \b^{sp} \del_k \b^{rk} \right) \nn\\
& +\tg_{uq} G^{qk} \iG_{sr} \del_p G^{rs} \del_k G^{up} + \tg_{lu} G^{up} \iG_{sr} \del_p G^{rs} \del_k G^{lk} + G^{qk} G^{up} \iG_{sr} \del_p G^{rs} \del_k \tg_{qu} \ ,\nn
\eea
\bea
\G^{pn}{}_k \G^{k}{}_{np} &= -\frac{1}{2} \del_k \tg_{ps} \del_m \tg_{uq} \left(\frac{1}{2} \tg^{us} \tg^{pq} \tg^{km} + \frac{1}{2} \tg^{us} \tg^{pq} \b^{lm} \tg_{lr} \b^{rk} - \tg^{sq} \b^{pm} \b^{uk} \right) \\
& + \b^{uk} \del_k \b^{qn} \del_n  \tg_{uq} + \frac{1}{2} \tg_{qs} \del_k \b^{qn} \del_n \b^{sk} \nn\\
& + \tg_{us} G^{sk} \iG_{lr} \del_k G^{rn} \del_n G^{ul} + \iG_{nl} \tg^{lq} \tg_{pu} \del_k G^{un} \del_m \tg_{qr} \left(G^{pm} G^{rk} - \tg^{rp} g^{km} \right) \nn\\
& + \frac{1}{2} \del_k G^{un} \del_m G^{rl} \left(-g^{km} \iG_{lu} \iG_{nr} + g_{nl} \tg_{pu} \tg_{rv} (G^{pm}G^{vk} -\tg^{pv} g^{km}) \right) \ ,\nn
\eea
\bea
g^{pn} \del_k \G^k{}_{np} &= \frac{1}{2} g^{km} \del_k (\tg^{pq} \del_m \tg_{pq}) + \left(\tg^{pk} \tg^{qn} - \b^{pk} \b^{qn} \right) \del_k \del_n \tg_{pq} - 2 \b^{qn} \tg_{pq} \del_k \del_n \b^{pk} \\
& + \del_k \tg_{su} \del_n \tg_{pq} \left(\tg^{uq} \left(\b^{sn} \b^{pk} - \frac{1}{2} g^{kn} \tg^{sp} - \tg^{sk} \tg^{pn} -\tg^{kp} \tg^{sn} \right) - \frac{1}{2} \tg^{pq} \left(\tg^{ks} \tg^{nu} + \b^{ns} \b^{uk} \right) \right) \nn\\
& - 2 \b^{pk} \del_k \tg_{pq} \del_n \b^{qn} + \frac{1}{2} \tg^{pq} \del_k \tg_{pq} \left( \tg_{ur} \b^{rl} \del_l \b^{uk} + \tg_{ur} \b^{rk} \del_l \b^{ul}\right) - \tg_{pq} \del_k \b^{pk} \del_n \b^{qn} \nn\\
& + \del_n G^{vl} \del_k G^{qn} \left(2 \iG_{lq} \tg_{vu} G^{uk} +\iG_{lv} \tg_{qp} G^{pk} \right) + \del_m G^{vn} \del_k G^{rk} \tg_{rs} G^{sm} \iG_{nv} \nn\\
& - \del_m G^{vl} \del_k G^{ps} \left(-g_{sl} \tg_{uv} G^{uk} \tg_{pq} G^{qm} + g^{km} \left(g_{sl} \tg_{pv} + 2 \iG_{sv} \iG_{lp} \right) \right) \nn\\
& + \del_m G^{vl} \del_k \tg_{pq} \left(2 G^{qm} \tg_{vu} G^{uk} \iG_{lr} \tg^{rp} + G^{pm} G^{qk} \iG_{lv} - 3 \delta^p_v g^{km} \iG_{lr} \tg^{rq} \right) \nn\\
& - g^{km} \iG_{vq} \tg^{qp} \del_m G^{vl} \del_k \tg_{lp} - g^{km} \iG_{nl} \left(\tg^{pn} \tg^{lu} \del_k \del_m \tg_{up} -\del_k \del_m \b^{ln} \right) \ .\nn
 \eea
One can check that these four terms reduce to their standard expressions given in \cite{allp11} when $\b=0$. Using (\ref{tR}) for ${\cal R}(\tg)$, one ends with 
\bea
\R(g) -\R(\tg) &= -  \left(\iG_{ru} \del_k G^{ur} + \frac{1}{2} \tg^{rs} \del_k \tg_{rs} \right) g^{km}
\left(\iG_{lv} \del_m G^{vl} + \frac{1}{2} \tg^{pn} \del_m \tg_{pn} \right) \label{R-R}\\
& + \left(2 \tg^{qp} \tg^{km}+ \tg^{pq} \b^{kr}\tg_{rs} \b^{ms} - \b^{pk} \b^{qm} \right) \del_k \del_m \tg_{pq} - 2 \b^{qm} \tg_{pq} \del_k \del_m \b^{pk} \nn\\
&  - 2 g^{km} \iG_{nl} \left(\tg^{pn} \tg^{lu} \del_k \del_m \tg_{up} -\del_k \del_m \b^{ln} \right) \nn\\
& +\del_k \tg_{su} \del_m \tg_{pq} \Bigg[\frac{1}{2}\tg^{uq} \left(\b^{sm} \b^{pk} - 4\tg^{sp} \tg^{km} -\tg^{kp} \tg^{sm} - \frac{5}{2} \tg^{sp} \b^{kr} \tg_{rv} \b^{mv} \right) \nn\\
& \qquad \qquad \qquad \qquad \qquad \qquad \qquad \qquad - \frac{1}{2} \tg^{pq} \left( 4\tg^{ks} \tg^{mu} +2 \b^{ms} \b^{uk} - \frac{1}{2} \tg^{km} \tg^{su}\right) \Bigg] \nn\\
& - 2 \b^{pk} \del_k \tg_{pq} \del_m \b^{qm} -\b^{uk} \del_m \tg_{uq} \del_k \b^{qm} + \tg^{pq} \del_k \tg_{pq} \left( \tg_{ur} \b^{rm} \del_m \b^{uk} + \tg_{ur} \b^{rk} \del_m \b^{um}\right) \nn\\
& - \tg_{pq} \left( \del_k \b^{pk} \del_m \b^{qm} +\frac{1}{2} \del_k \b^{qm} \del_m \b^{pk} \right) \nn\\
& + 2 \del_m G^{vn} \del_k G^{rk} \tg_{rs} G^{sm} \iG_{nv} + \del_m G^{vl} \del_k G^{qm} \left(\iG_{lq} \tg_{vu} G^{uk} + 2 \iG_{lv} \tg_{qp} G^{pk} \right)  \nn\\
& - \del_m G^{vl} \del_k G^{ps} \left(- \frac{1}{2} g_{sl} \tg_{uv} G^{uk} \tg_{pq} G^{qm} +\frac{1}{2} g^{km} \left(g_{sl} \tg_{pv} + 5 \iG_{sv} \iG_{lp} \right) \right) \nn\\
& + \del_m G^{vl} \del_k \tg_{pq} \left( G^{qm} \tg_{vu} G^{uk} \iG_{lr} \tg^{rp} +2 G^{pm} G^{qk} \iG_{lv} - 3 \delta^p_v g^{km} \iG_{lr} \tg^{rq} \right) \nn\\
& - 2 g^{km} \iG_{vq} \tg^{qp} \del_m G^{vl} \del_k \tg_{lp} \ .\nn
\eea
One can check that this vanishes for $\b=0$. Plugging the assumption of \cite{allp11} in the previous expression, we also recover the formula given there. For the sake of brevity or later convenience, we have left a few terms containing $g_{km}$ and $g^{km}$.

\subsection*{Dilaton terms, second order derivative terms, and total derivative}

Using the definition  \eqref{eq:dilintro} of the new dilaton $\tp$, we showed in \cite{allp11} that
\bea
\partial_m \phi &= \partial_m \tp -\frac{1}{2} A_m \ , \label{diffphiAm}\\
4 \left((\partial \phi)^2- (\partial \tp)^2\right) &= 4 (g^{km}-\tg^{km}) \del_k \tp \del_m \tp  + g^{km} A_k A_m - 4 g^{km} A_k \del_m \tp  \ ,\label{dildiff}
\eea
where we mean $(\partial \phi)^2=g^{km} \del_k \phi \del_m \phi$, $(\partial \tp)^2=\tg^{km} \del_k \tp \del_m \tp$, and we introduced for convenience
\beq
A_m=\iG_{kl}\partial_m\b^{lk} + \iG_{kl}\tg^{ln}\partial_m\tg_{np} \b^{pk}\ . \label{defA}
\eeq
One can also show that $A_m=\tg^{pq} \del_m \tg_{pq}+ \iG_{lk} \del_m G^{kl}$. Then the first row in \eqref{R-R} becomes
\bea
& -\left(\iG_{ru} \del_k G^{ur} + \frac{1}{2} \tg^{rs} \del_k \tg_{rs} \right) g^{km} \left(\iG_{lv} \del_m G^{vl} + \frac{1}{2} \tg^{pn} \del_m \tg_{pn} \right)
= -g^{km} A_k A_m \label{eqbla}\\
& \qquad\qquad\qquad\qquad\qquad\qquad + g^{km} \tg^{pq} \iG_{ln} \del_k G^{nl} \del_m \tg_{pq} + \frac{3}{4} g^{km} \tg^{pq} \tg^{uv} \del_k \tg_{pq} \del_m \tg_{uv} \ .\nn
\eea
Adding to \eqref{R-R} the dilaton terms \eqref{dildiff}, one cancels the term in $g^{km} A_k A_m$ using \eqref{eqbla}. 

We now turn to the second order derivative terms, contained in the second and third row of \eqref{R-R}. These terms cannot be canceled against any of the remaining terms in $\L$, which have only first order derivatives. So we rewrite them with a total derivative term. For combinations of fields $f$ and $F^{km}$ one has generically
\beq
F^{km} \del_k \del_m f = \ \frac{\del_k\left(e^{-2d}\ F^{km} \del_m f \right)}{e^{-2d}} + \Bigg( \left( 2\del_k \tp -\frac{1}{2} \tg^{pq} \del_k \tg_{pq} \right) F^{km} -\del_k F^{km} \Bigg) \del_m f \ ,\label{intpart}
\eeq
where we used \eqref{eq:dilintro} for the measure. Before using this formula, let us rewrite slightly the terms of interest in \eqref{R-R} using the definition of $G$. One has
\bea
& \left(2 \tg^{qp} \tg^{km}+ \tg^{pq} \b^{kr}\tg_{rs} \b^{ms} - \b^{pk} \b^{qm} \right) \del_k \del_m \tg_{pq} - 2 \b^{qm} \tg_{pq} \del_k \del_m \b^{pk} \label{secondorderterms} \\
&  - 2 g^{km} \iG_{nl} \left(\tg^{pn} \tg^{lu} \del_k \del_m \tg_{up} -\del_k \del_m \b^{ln} \right) \nn\\
&= \left(2 \tg^{qp} \tg^{km}+ \tg^{pq} \b^{kr}\tg_{rs} \b^{ms} - \b^{pk} \b^{qm} \right) \del_k \del_m \tg_{pq} - 2 \b^{qm} \tg_{pq} \del_k \del_m \b^{pk} - 2 g^{km} \tg^{pq} \del_k \del_m \tg_{pq} \nn \\
&  + 2 g^{km} \iG_{nl} \left(\b^{pn} \tg^{lu} \del_k \del_m \tg_{up} +\del_k \del_m \b^{ln} \right) \nn
\eea
Now using \eqref{intpart}, one obtains for the last two terms of \eqref{secondorderterms}
\bea
& 2 g^{km} \iG_{nl} \left(\b^{pn} \tg^{lu} \del_k \del_m \tg_{up} +\del_k \del_m \b^{ln} \right) - e^{2d}\ \del_k \left(\dots \right) \label{intbp1}\\
& =  4 g^{km} A_k \del_m \tp -g^{km} \tg^{pq} \iG_{ln} \del_k G^{nl} \del_m \tg_{pq} \nn\\
& - g^{km} \tg^{pq} \tg^{uv} \del_k \tg_{pq} \del_m \tg_{uv} + 2 g^{km} \tg^{pr} \tg^{us} \del_m \tg_{up} \del_k \tg_{rs} \nn\\
& -2 \tg^{pq} \del_m \tg_{pq} \left(\tg_{pr} \b^{rk} \del_k \b^{pm} + \tg_{pr} \b^{rm} \del_k \b^{pk} - \del_k \tg_{uv} (\tg^{ku} \tg^{vm} + \b^{mu} \b^{vk}) \right) \nn\\
& -2 \tg_{rs} G^{rk} \iG_{lp} \del_m G^{pl} \del_k G^{sm} - 2 \tg_{rs} G^{rm} \iG_{lp} \del_m G^{pl} \del_k G^{sk} + 2 g^{km} \iG_{sl} \iG_{nr} \del_m G^{ln} \del_k G^{rs} \nn\\
& -2 G^{pm} G^{uk} \iG_{ns} \del_k G^{sn} \del_m \tg_{up} + 2 g^{km} \iG_{nl} \del_k (\tg^{lu} \tg^{pn}) \del_m \tg_{up} \ ,\nn
\eea
where we used the definition of $G$, and where the total derivative is given by (see \eqref{defA})
\beq
\del_k \left(e^{-2\tp} \sqrt{|\tg|}\  2 g^{km} \iG_{nl} \left(\b^{pn} \tg^{lu} \del_m \tg_{up} +\del_m \b^{ln} \right) \right) = \del_k \left(e^{-2\tp} \sqrt{|\tg|}\  2 g^{km} A_m \right) \ .\label{totder1}
\eeq
The other second order derivative terms in \eqref{secondorderterms} need more attention. A first use of \eqref{intpart} gives the total derivative $\del_k \left(e^{-2d}\ \left( (\tg^{pq} (\tg^{km}-g^{km} ) - \b^{pk} \b^{qm} ) \del_m \tg_{pq} - 2 \b^{qm} \tg_{pq} \del_m \b^{pk} \right) \right)$, where the first equality in \eqref{ghat} was used. However, a piece of this total derivative, namely $\del_k \left(e^{-2d}\ \left(-\b^{qm} \tg_{pq} \del_m \b^{pk} + \b^{pk} \tg_{pq} \del_m \b^{qm} \right) \right)$, can be developed. Indeed, this piece has the particularity of producing no second order derivative terms. Doing so, one is left with a simpler total derivative, given below. These manipulations finally result in 
\bea
& \left(2 \tg^{qp} \tg^{km}+ \tg^{pq} \b^{kr}\tg_{rs} \b^{ms} - \b^{pk} \b^{qm} \right) \del_k \del_m \tg_{pq} - 2 \b^{qm} \tg_{pq} \del_k \del_m \b^{pk} -2 g^{km} \tg^{pq} \del_k \del_m \tg_{pq} \label{intbp2} \\
&= -2 \del_k \tp \left(\tg^{pq} \del_m \tg_{pq} \b^{kr}\tg_{rs} \b^{ms} + \del_m (\b^{rk}\tg_{rs} \b^{sm})  \right) - \b^{kr}\tg_{rs} \b^{ms} \tg^{qu} \tg^{vp} \del_m \tg_{pq} \del_k \tg_{uv} \nn\\
& +\frac{1}{2} \tg^{pq} \del_k \tg_{pq} \left(\tg^{rs} \del_m \tg_{rs} \b^{ku}\tg_{uv} \b^{mv} + 3 \tg_{rs} \b^{sm} \del_m \b^{rk}
+ 3 \tg_{rs} \b^{rk} \del_m \b^{sm} + 3 \b^{rk} \b^{sm} \del_m \tg_{rs} \right) \nn\\
& +2 \del_m \tg_{pq} \left(\b^{qm} \del_k \b^{pk} + \b^{pk} \del_k \b^{qm} \right) + \tg_{pq} \left( \del_k \b^{qm} \del_m \b^{pk} + \del_k \b^{qk} \del_m \b^{pm} \right) \nn\\
& +e^{2d} \ \del_k \left(e^{-2d}\ \left( \tg^{pq} (\tg^{km}-g^{km} ) \del_m \tg_{pq} - \del_m (g^{km}-\tg^{km}) \right) \right) \ . \label{totder2}
\eea
The two total derivative terms \eqref{totder1} and \eqref{totder2}  actually combine nicely. Indeed, using \eqref{diffphiAm}, \eqref{eq:dilintro}, and then \eqref{lndet}, \eqref{eq:lndet}, one gets $2 A_m= \tg^{pq} \del_m \tg_{pq} - g^{pq} \del_m g_{pq}$. Then, the sum of the total derivatives in \eqref{totder1} and \eqref{totder2} becomes
\beq
\del_k \left[e^{-2\tp} \sqrt{|\tg|}\ \left( \tg^{km} \tg^{pq} \del_m \tg_{pq} - g^{km} g^{pq} \del_m g_{pq} - \del_m ( g^{km}- \tg^{km} ) \right) \right] \ .\label{totder}
\eeq
It is illuminating to compare this total derivative with that in \eqref{eq:Ldftcompare}, which is obtained when rewriting the DFT Lagrangian in a form that contains the Ricci scalar. Taking the difference between the DFT total derivatives for $g$ and $\tg$, as suggested in \eqref{diagr}, we reproduce \eqref{totder}.

Putting all pieces together, namely \eqref{R-R}, \eqref{dildiff}, \eqref{eqbla}, \eqref{secondorderterms}, \eqref{intbp1}, \eqref{intbp2} and \eqref{totder}, nice cancelations occur. We finally obtain (with the total derivative given by \eqref{totder})
\bea
& \!\!\!\!\! \R(g) -\R(\tg) + 4 \left((\partial \phi)^2- (\partial \tp)^2\right) - e^{2d}\ \del_k(\dots) \label{result0}\\
&= 4 (g^{km}-\tg^{km}) \del_k \tp \del_m \tp -2 \del_k \tp \left(\tg^{pq} \del_m \tg_{pq} (g^{km}- \tg^{km}) + \del_m (g^{km}- \tg^{km})  \right) \nn  \\
& +\frac{1}{2} \del_k \tg_{su} \del_m \tg_{pq} \Bigg[\tg^{uq} \left(\b^{sm} \b^{pk}  -\tg^{kp} \tg^{sm} - \frac{1}{2} \tg^{sp} \b^{kr} \tg_{rv} \b^{mv} \right)
  + \tg^{pq} \left(\b^{sm} \b^{uk} + \frac{1}{2} \tg^{su} (g^{km} - \tg^{km} ) \right) \Bigg] \nn\\
& +\b^{uk} \del_m \tg_{uq} \del_k \b^{qm} + \frac{1}{2} \tg^{pq} \del_k \tg_{pq} \left( \tg_{ur} \b^{rm} \del_m \b^{uk} + \tg_{ur} \b^{rk} \del_m \b^{um}\right)
 + \frac{1}{2} \tg_{pq} \del_k \b^{qm} \del_m \b^{pk} \nn\\
&  - \frac{1}{2} \del_m G^{vl} \del_k G^{ps} \left(- g_{sl} \tg_{uv} G^{uk} \tg_{pq} G^{qm} + g^{km} \left(g_{sl} \tg_{pv} + \iG_{sv} \iG_{lp} \right) \right) \nn\\
& + \iG_{lq} \tg_{vu} G^{uk} \del_m G^{vl} \del_k G^{qm} + \del_m G^{vl} \del_k \tg_{pq} \left( G^{qm} \tg_{vu} G^{uk} \iG_{lr} \tg^{rp}  - 3 \delta^p_v g^{km} \iG_{lr} \tg^{rq} \right) \nn\\
& - 2 g^{km} \iG_{vq} \tg^{qp} \del_m G^{vl} \del_k \tg_{lp} + 2 g^{km} \iG_{nl} \del_k (\tg^{lu} \tg^{pn}) \del_m \tg_{up} \ .\nn
\eea

Let us now rearrange and simplify a bit \eqref{result0}. First, thanks to the symmetry of $(n,l)$ in the very last term of \eqref{result0}, it can be rewritten as $-4 g^{km} \iG_{vq} \tg_{lp} \del_k \tg^{qp} \del_m \tg^{vl}$. Using this, one can show that the last row of \eqref{result0} cancels with $2$ out of $3$ of the term in $\delta^p_v$, in the last but one row. Secondly, we pick the following terms from \eqref{result0}
\bea
& 4 (g^{km}-\tg^{km}) \del_k \tp \del_m \tp -2 \del_k \tp \left(\tg^{pq} \del_m \tg_{pq} (g^{km}- \tg^{km}) + \del_m (g^{km}- \tg^{km})  \right) \label{arrange1}\\
& + \frac{1}{2} \tg^{pq} \del_k \tg_{pq} \Bigg[\del_m \tg_{su} \left(\b^{sm} \b^{uk} + \frac{1}{2} \tg^{su} (g^{km} - \tg^{km} ) \right) + \tg_{ur} \b^{rm} \del_m \b^{uk} + \tg_{ur} \b^{rk} \del_m \b^{um} \Bigg]\ . \nn
\eea
Using the first equality in \eqref{ghat}, these terms can be recombined into an expression proportional to $2 \del_k \tp - \frac{1}{2} \tg^{pq} \del_k \tg_{pq} = - \del_k \ln \left( e^{-2\tp} \sqrt{|\tg|} \right)$.

Given these rearrangings, \eqref{result0} eventually simplifies to
\bea
& \!\!\!\!\! \R(g) -\R(\tg) + 4 \left((\partial \phi)^2- (\partial \tp)^2\right) - e^{2d}\ \del_k(\dots) \label{Rphidgen}\\
&= (g^{km} - \tg^{km} ) \del_k \ln\left( e^{-2\tp} \sqrt{|\tg|} \right) \del_m \ln\left( e^{-2\tp} \sqrt{|\tg|} \right) + \del_m (g^{km} - \tg^{km}) \del_k \ln\left( e^{-2\tp} \sqrt{|\tg|} \right) \nn  \\
& + \frac{1}{2}\tg^{uq} \del_k \tg_{su} \del_m \tg_{pq} \left(\b^{sm} \b^{pk}  -\tg^{kp} \tg^{sm} - \frac{1}{2} \tg^{sp} \b^{kr} \tg_{rv} \b^{mv} \right)  +  \del_k \b^{qm} \left (\b^{uk} \del_m \tg_{uq}   + \frac{1}{2} \tg_{pq} \del_m \b^{pk} \right) \nn\\
& - \frac{1}{2} \del_m G^{vl} \del_k G^{ps} \left(- g_{sl} \tg_{uv} G^{uk} \tg_{pq} G^{qm} + g^{km} \left(g_{sl} \tg_{pv} + \iG_{sv} \iG_{lp} \right) \right) \nn\\
&  + \iG_{lq} \tg_{vu} G^{uk} \del_m G^{vl} \del_k G^{qm} + \del_m G^{vl} \del_k \tg_{pq} \left( G^{qm} \tg_{vu} G^{uk} \iG_{lr} \tg^{rp}  - \delta^p_v g^{km} \iG_{lr} \tg^{rq} \right) \ .\nn
\eea

\subsection*{The $H$  flux term}

In \cite{allp11}, it was shown that 
\beq
\frac{1}{3}H_{ijk}H^{ijk}=3 \left((I) + (II) + (III)\right) \ , \label{pH2}
\eeq
\bea
{\rm with}\qquad 3 (I)=&\left(
\tg_{p_1 p_2} \tg_{q_1 q_2} \tg_{s_1 s_2}-
\tg_{p_1 s_2} \tg_{q_1 q_2} \tg_{s_1 p_2}-
\tg_{p_1 p_2} \tg_{q_1 s_2} \tg_{s_1 q_2}
\right)
D_{\epsilon}^{s_1} \b^{p_1 q_1} D_{\epsilon}^{s_2} \b^{p_2 q_2} \label{finalI}\\
3 (II)=&4\left(
\tg_{p_1 p_2} \tg_{t_1 q_2} \tg_{s_1 s_2}-
\tg_{p_1 s_2} \tg_{t_1 q_2} \tg_{s_1 p_2}-
\tg_{p_1 p_2} \tg_{t_1 s_2} \tg_{s_1 q_2}
\right) \label{finalII} \\
& \times \b^{t_1 t_2} (G_{\epsilon}^{-1})_{q_1 t_2} 
D_{\epsilon}^{s_1} G_{\epsilon}^{p_1 q_1} D_{\epsilon}^{s_2} \b^{p_2 q_2} \nn\\
3 (III)=&2 (
\tg_{p_1 p_2} \tg_{t_1 t_2} \tg_{s_1 s_2}-
\tg_{p_1 s_2} \tg_{t_1 t_2} \tg_{s_1 p_2}-
\tg_{p_1 p_2} \tg_{t_2 s_2} \tg_{s_1 t_1} \label{finalIII}\\
&-\tg_{p_1 t_1} \tg_{p_2 t_2} \tg_{s_1 s_2}+
\tg_{p_1 s_2} \tg_{t_2 p_2} \tg_{s_1 t_1}+
\tg_{p_1 t_1} \tg_{t_2 s_2} \tg_{s_1 p_2}) \nn\\
&\times\left(\delta_{q_1}^{t_2} - (G_{\epsilon}^{-1})_{q_1 u_2} \tg^{u_2 t_2}\right)
\left(\delta_{q_2}^{t_1} - (G_{\epsilon}^{-1})_{q_2 u_1} \tg^{u_1 t_1}\right) 
D_{\epsilon}^{s_1} G_{\epsilon}^{p_1 q_1} D_{\epsilon}^{s_2} G_{\epsilon}^{p_2 q_2} \ ,\nn
\eea
where $\epsilon=\pm1$ was left unspecified and the notation $D_{\epsilon}^p =G_{\epsilon}^{p q} \partial_q $ was introduced. Here we develop and rewrite these expressions further.

Let us first note that $\delta_{q_1}^{t_2} - (G_{\epsilon}^{-1})_{q_1 u_2} \tg^{u_2 t_2}= (G_{\epsilon}^{-1})_{q_1 u_2}\ \epsilon \b^{u_2 t_2}$. Applying this to the term of $3(III)$ in $\delta_{q_2}^{t_1}$, and using the antisymmetry appearing between $p_2$ and $q_2$, this term reduces to $-3(II)$. Therefore, we get that
\bea
3(II)+3(III) &=  -2 (G_{\epsilon}^{-1})_{q_2 u_1} \tg^{u_1 t_1} \left(\delta_{q_1}^{t_2} - (G_{\epsilon}^{-1})_{q_1 u_2} \tg^{u_2 t_2}\right)
D_{\epsilon}^{s_1} G_{\epsilon}^{p_1 q_1} D_{\epsilon}^{s_2} G_{\epsilon}^{p_2 q_2} \\
&\times(\tg_{p_1 p_2} \tg_{t_1 t_2} \tg_{s_1 s_2}-
\tg_{p_1 s_2} \tg_{t_1 t_2} \tg_{s_1 p_2}-
\tg_{p_1 p_2} \tg_{t_2 s_2} \tg_{s_1 t_1} \nn\\
&-\tg_{p_1 t_1} \tg_{p_2 t_2} \tg_{s_1 s_2}+
\tg_{p_1 s_2} \tg_{t_2 p_2} \tg_{s_1 t_1}+
\tg_{p_1 t_1} \tg_{t_2 s_2} \tg_{s_1 p_2}) \nn
\eea
We then multiply $\tg^{u_1 t_1} \left(\delta_{q_1}^{t_2} - (G_{\epsilon}^{-1})_{q_1 u_2} \tg^{u_2 t_2}\right)$ with the parentheses containing metrics, use again a few symmetry arguments and finally obtain
\bea
3(II)+3(III) &=  -4 \epsilon \tg_{pq} \tg_{nr} G_{\epsilon}^{rm} \del_k \b^{np} \del_m G_{\epsilon}^{qk} - 2 \tg_{np} \del_k G_{\epsilon}^{nm} \del_m G_{\epsilon}^{pk} \label{eq23}\\
& \!\!\!\!\!\!\!\!\!\!\!\!\!\!\! +4 \del_k \tg^{np} \del_m G_{\epsilon}^{qr} (G_{\epsilon}^{-1})_{rp} (\tg_{nq} g^{km} - \tg_{ns} G_{\epsilon}^{sm} \tg_{ql} G_{\epsilon}^{lk}) + 4 \del_k G_{\epsilon}^{np} \del_m G_{\epsilon}^{qk} (G_{\epsilon}^{-1})_{pq} \tg_{nr} G_{\epsilon}^{rm} \nn\\
& \!\!\!\!\!\!\!\!\!\!\!\!\!\!\! + 2 \del_k G_{\epsilon}^{np} \del_m G_{\epsilon}^{qr} \left[ (g_{pr}-2 (G_{\epsilon}^{-1})_{rp}) (\tg_{nq} g^{km} - \tg_{ns} G_{\epsilon}^{sm} \tg_{ql} G_{\epsilon}^{lk}) -g^{km} (G_{\epsilon}^{-1})_{pq} (G_{\epsilon}^{-1})_{rn}  \right] \nn
\eea
The quantities multiplying $(g_{pr}-2 (G_{\epsilon}^{-1})_{rp})$ are actually symmetric in $(r,p)$, so we can use that $\forall \epsilon\ , \ (G_{\epsilon}^{-1})_{(rp)} = \frac{1}{2} \left(\iG_{rp} + \iG_{pr}\right) = g_{rp}$, as can be seen in \eqref{ghat}. In addition, we develop the first line of \eqref{eq23}, so we finally obtain
\bea
3(II)+3(III) \\
& \!\!\!\!\!\!\!\!\!\!\!\!\!\!\! = -2 \tg_{pq} \left(\del_k \tg^{pm} \del_m \tg^{qk} + \del_k \b^{mp} \del_m \b^{qk} +2 \tg_{nr} \b^{rm} \del_k \b^{np} (\del_m \tg^{qk} +\epsilon \del_m \b^{qk} )  \right) \nn\\
& \!\!\!\!\!\!\!\!\!\!\!\!\!\!\! +4 \del_k \tg^{np} \del_m G_{\epsilon}^{qr} (G_{\epsilon}^{-1})_{rp} (\tg_{nq} g^{km} - \tg_{ns} G_{\epsilon}^{sm} \tg_{ql} G_{\epsilon}^{lk}) + 4 \del_k G_{\epsilon}^{np} \del_m G_{\epsilon}^{qk} (G_{\epsilon}^{-1})_{pq} \tg_{nr} G_{\epsilon}^{rm} \nn\\
& \!\!\!\!\!\!\!\!\!\!\!\!\!\!\! + 2 \del_k G_{\epsilon}^{np} \del_m G_{\epsilon}^{qr} \left[ -g_{pr} (\tg_{nq} g^{km} - \tg_{ns} G_{\epsilon}^{sm} \tg_{ql} G_{\epsilon}^{lk}) -g^{km} (G_{\epsilon}^{-1})_{pq} (G_{\epsilon}^{-1})_{rn}  \right]\ . \nn
\eea
From \eqref{finalI}, we develop and get, using some symmetry arguments
\bea
3(I) &= \tg_{pr} \tg_{nq} g^{km} \del_k \b^{np} \del_m \b^{qr} 
-2 \tg_{pr} \tg_{ns} \b^{sm} \tg_{ql} \b^{lk} \del_k \b^{np} \del_m \b^{qr}\\
& -2 \tg_{pr} \left(\del_k \b^{mp} \del_m \b^{kr} + 2\epsilon \tg_{qs} \b^{sk} \del_k \b^{mp} \del_m \b^{qr}  \right) \nn \\
& = \tg_{pr} \tg_{nq} \tg^{km} \del_k \b^{np} \del_m \b^{qr} +2 \b^{lk} \b^{sm} \del_k \b^{rn} \del_m \b^{pq} \left(\frac{1}{2} \tg_{qn} \tg_{pr} \tg_{sl} - \tg_{qn} \tg_{pl} \tg_{rs} \right) \nn \\
& -2 \tg_{pr} \left(\del_k \b^{mp} \del_m \b^{kr} + 2\epsilon \tg_{qs} \b^{sk} \del_k \b^{mp} \del_m \b^{qr}  \right) \nn \, .
\eea

Finally, combining all these results as in \eqref{pH2}, we obtain
\bea
-\frac{1}{12} H_{ijk}H^{ijk} & = -\frac{1}{4} \tg_{pr} \tg_{nq} \tg^{km} \del_k \b^{np} \del_m \b^{qr} -\frac{1}{2} \b^{lk} \b^{sm} \del_k \b^{rn} \del_m \b^{pq} \left(\frac{1}{2} \tg_{qn} \tg_{pr} \tg_{sl} - \tg_{qn} \tg_{pl} \tg_{rs} \right) \nn\\
& +\frac{1}{2} \tg_{pq} \del_k \tg^{pm} \del_m \tg^{qk} +  \tg_{pq} \tg_{nr} \b^{rm} \del_k \b^{np} \del_m \tg^{qk} \label{H2}\\
& - \del_k \tg^{np} \del_m G_{\epsilon}^{qr} (G_{\epsilon}^{-1})_{rp} (\tg_{nq} g^{km} - \tg_{ns} G_{\epsilon}^{sm} \tg_{ql} G_{\epsilon}^{lk}) - \del_k G_{\epsilon}^{np} \del_m G_{\epsilon}^{qk} (G_{\epsilon}^{-1})_{pq} \tg_{nr} G_{\epsilon}^{rm} \nn\\
& -\frac{1}{2} \del_k G_{\epsilon}^{np} \del_m G_{\epsilon}^{qr} \left[ -g_{pr} (\tg_{nq} g^{km} - \tg_{ns} G_{\epsilon}^{sm} \tg_{ql} G_{\epsilon}^{lk}) -g^{km} (G_{\epsilon}^{-1})_{pq} (G_{\epsilon}^{-1})_{rn}  \right] \ ,\nn
\eea
and in the following we choose for the free parameter $\epsilon=1$.

\subsection*{Combining results}

Combining \eqref{Rphidgen} and \eqref{H2}, and using $e^{-2d} = e^{-2\tp} \sqrt{|\tg|}$, we finally obtain the following equality, where the total derivative is given in \eqref{totder}
\bea
&\R(g) -\R(\tg) + 4 \left((\partial \phi)^2- (\partial \tp)^2\right)- e^{2d}\del_k(\dots) 
-\frac{1}{12} H_{ijk}H^{ijk}
 \label{final2} \\
& = 4(g^{km} - \tg^{km} )\ \del_k d\ \del_m d -2 \del_m (g^{km} - \tg^{km})\ \del_k d \nn \\
& -\frac{1}{4} \tg_{pr} \tg_{nq} \tg^{km} \del_k \b^{np} \del_m \b^{qr} + \frac{1}{2} \tg_{pq} \del_k \b^{qm} \del_m \b^{pk} \nn\\
& +\b^{uk} \del_m \tg_{uq} \del_k \b^{qm} - \tg^{qk} \tg_{nr} \b^{rm} \del_k \b^{np} \del_m \tg_{pq} \nn\\
& + \frac{1}{2}\tg^{uq} \del_k \tg_{su} \del_m \tg_{pq}  \left(\b^{sm} \b^{pk}  - \frac{1}{2} \tg^{sp} \b^{kr} \tg_{rv} \b^{mv} \right) \nn\\
& -\frac{1}{2} \b^{lk} \b^{sm} \del_k \b^{rn} \del_m \b^{pq} \left(\frac{1}{2} \tg_{qn} \tg_{pr} \tg_{sl} - \tg_{qn} \tg_{pl} \tg_{rs} \right) \nn\ ,
\eea
It is remarkable that all $\iG$ have been cancelled. In order to match with \eqref{eq:Lfinal}, note that the last row of \eqref{final2} gives the $R$ flux term, and $g^{km}-\tg^{km}=\beta^{kp} \tg_{pq} \beta^{mq}$ follows from \eqref{ghat}.

\end{appendix}


\newpage

\providecommand{\href}[2]{#2}\begingroup\raggedright\endgroup

\end{document}